\PassOptionsToPackage{table}{xcolor} %
\documentclass[journal,10pt]{IEEEtran}
\usepackage{orcidlink}

\AtBeginDocument{%
  \providecommand\BibTeX{{%
    \normalfont B\kern-0.5em{\scshape i\kern-0.25em b}\kern-0.8em\TeX}}}

\usepackage[T1]{fontenc} %

\usepackage{hyperref}
\usepackage{pdflscape}
\usepackage{makecell}
\usepackage{booktabs}
\usepackage[para, flushleft]{threeparttable}
\usepackage{tabularx}
\usepackage{colortbl}
\newcolumntype{L}[1]{>{\raggedright\arraybackslash}p{#1}} %
\newcolumntype{C}[1]{>{\centering\arraybackslash}p{#1}} %
\newcolumntype{R}[1]{>{\raggedleft\arraybackslash}p{#1}} %

\usepackage{subcaption}

\usepackage{csquotes} %

\DeclareQuoteStyle[american]{english} %
        {\itshape\textquotedblleft}
        [\textquotedblleft]
        {\textquotedblright}
        [0.05em]
        {\textquoteleft}
        {\textquoteright}

\usepackage{tikz}
\usetikzlibrary{shapes,arrows, arrows.meta,matrix, positioning, fit, calc, decorations.pathreplacing}

\newbool{displaycomments}
\setbool{displaycomments}{false}

\ifbool{displaycomments}{%
    \definecolor{springgreen}{rgb}{0.0, 1.0, 0.5}
    \definecolor{dandelion}{HTML}{f0e130}
    \definecolor{navy}{HTML}{002F6C}
    \newcommand\COMMENTX[2]{{\bf({#1}: {#2})}}
    \newcommand\sascha[1]{\COMMENTX{sascha}{{{\color{orange}#1}}}}
    \newcommand\yas[1]{\COMMENTX{yas}{{{\color{pink}#1}}}}
    \newcommand\jan[1]{\COMMENTX{jan}{{{\color{red}#1}}}}
    \newcommand\philip[1]{\COMMENTX{philip}{{{\color{violet}#1}}}}
    \newcommand\niklas[1]{\COMMENTX{niklas}{{{\color{navy}#1}}}}
    \newcommand\supervisor[1]{\COMMENTX{for supervisor}{{{\color{purple}#1}}}}
    \newcommand\todo[1]{\COMMENTX{todo}{{{\color{red}#1}}}}
    \newcommand\draft[1]{{{\color{gray}#1}}}
    \newcommand\red[1]{{{\color{red}#1}}}
}{%
    \newcommand\sascha[1]{}
    \newcommand\chris[1]{}
    \newcommand\yas[1]{}
    \newcommand\jan[1]{}
    \newcommand\philip[1]{}
    \newcommand\niklas[1]{}
    \newcommand\supervisor[1]{}
    \newcommand\todo[1]{}
    \newcommand\draft[1]{}
    \newcommand\red[1]{}
}

\newcommand{\sequestcite}[4]{{\blockquote[\href{#2}{#1}]{#4}}}

\usepackage{xfp}
\newcommand{\printpercent}[2]{\fpeval{round(#1*100/#2,1)}\%}
\newcommand{\printstats}[1]{#1, \printpercent{#1}{\var{posts.started_with_duplicates}}}
\newcommand{\printstatsc}[2]{#1}
\newcommand{\printpercentoverall}[1]{\printpercent{#1}{\var{posts.started_with_duplicates}}}

\newcommand{\boldparagraph}[1]{\paragraph{#1}}
\makeatletter
\renewcommand\boldparagraph{\@startsection{paragraph}{4}{0\parindent}%
    {0.6ex plus 0.6ex minus 0.2ex}%
    {0ex}%
    {\normalfont\normalsize\bfseries\maybe@addperiod}*%
}
\newcommand{\maybe@addperiod}[1]{%
    \let\@period\@empty%
    \def\@IEEEsectpunct{}%
    #1\@addpunct{.}\enspace%
}
\makeatother

\usepackage[usetwoside=true]{mdframed}
\usepackage{tcolorbox}
\tcbuselibrary{breakable}
\tcbuselibrary{skins}
\definecolor{darkgray}{gray}{0.3}
\tcbset{}
\newtcolorbox{summaryBox}[2][]
{
    enhanced,
    breakable,
    frame hidden,
    borderline west = {3pt}{0pt}{lightgray},
    colback         = white,
    size            = fbox,
    left            = 0.3em,
    enlarge top by  = 0.5em,%
    coltitle        = black,
    title           = {\color{darkgray} \textbf{#2}. },
    attach title to upper,
    #1,
}

\newlength\bubblesize
\newcommand{\ie}{i.\,e.,}
\newcommand{\eg}{e.\,g.,}

\newcommand\definevar[2]{%
  \expandafter\newcommand\csname var#1var\endcsname{#2}%
}
\newcommand{\var}[1]{\ifcsname var#1var\endcsname%
        \csname var#1var\endcsname%
    \else\PackageWarning{Var}{`#1' does not exist.}%
        \red{TODO}%
    \fi%
}

\usepackage{xurl}
\PassOptionsToPackage{hyphens,spaces,obeyspaces}{xurl}

\usepackage[
    backend=biber,
    isbn=false,
    doi=false,
    url=false,
    eprint=false,
    defernumbers=true,
    mincitenames=1,
    maxcitenames=1,
    uniquelist=false,
    datamodel=acmdatamodel,
   style=acmnumeric,
]{biblatex}

\DeclareCiteCommand{\citeauthorfull}
    {\boolfalse{citetracker}%
    \boolfalse{pagetracker}%
    \DeclareNameAlias{labelname}{given-family}%
    \usebibmacro{prenote}}
    {\ifciteindex
        {\indexnames{labelname}}
        {}%
        \printnames{labelname}}
        {\multicitedelim}
{\usebibmacro{postnote}}

\definevar{posts.started}{309}
\definevar{posts.started_with_duplicates}{316}
\definevar{posts.allever}{830}
\definevar{posts.allever_tech_sites}{694 \todo{update}}
\definevar{posts.answered.accepted}{96}
\definevar{posts.answered}{194}
\definevar{posts.unanswered}{122}
\definevar{posts.negative_voting}{23}
\definevar{posts.closed}{24}
\definevar{posts.needsmorefocus}{6}
\definevar{posts.notsuitable}{5}
\definevar{posts.clarity.and.opinionbased}{3}

\definevar{questions.type.category.count}{6}
\definevar{questions.type.howto}{30}
\definevar{questions.type.troubleshooting}{15}
\definevar{questions.type.unexpected}{8}
\definevar{questions.type.benchmarks}{19}
\definevar{questions.type.conceptual}{32}
\definevar{questions.type.evaluation}{9}

\definevar{questions.count.duplicates}{7}
\definevar{questions.count.duplicates.word}{seven}

\definevar{questions.sampled.sample}{96}

\definevar{questions.unrelated.sample}{13}
\definevar{questions.closed.sample}{3}
\definevar{questions.duplicate.sample}{2}
\definevar{questions.negative_vote_count.sample}{8}
\definevar{questions.skipped.sample}{24}

\definevar{questions.closed_re_additions.sample}{1}
\definevar{questions.duplicate_re_additions.sample}{2}
\definevar{questions.negative_vote_count_re_additions.sample}{3}
\definevar{questions.total_re_additions.sample}{6}

\definevar{questions.duplicate_additions.sample}{2}
\definevar{questions.skipped_after_analysis.sample}{18}

\definevar{questions.included_with_additions.sample}{80}
\definevar{questions.sampled_with_additions.sample}{98}

\definevar{questions.remaining}{220}
\definevar{questions.duplicate.remaining}{4}

\definevar{challenge.count}{7}
\definevar{challenge.count.word}{seven}

\definevar{challenge.knowledge.count}{67}
\definevar{challenge.knowledge.domain.os.count}{40}
\definevar{challenge.knowledge.securityaspect.deployment.count}{21}
\definevar{challenge.knowledge.securityaspect.accesscontrol.count}{20}
\definevar{challenge.knowledge.driver.notspecified.count}{15}
\definevar{challenge.knowledge.driver.attack.count}{29}

\definevar{challenge.writingconfig.count}{79}
\definevar{challenge.writingconfig.domain.os.count}{35}
\definevar{challenge.writingconfig.domain.serverapp.count}{34}
\definevar{challenge.writingconfig.securityaspect.deployment.count}{34}
\definevar{challenge.writingconfig.securityaspect.accesscontrol.count}{30}
\definevar{challenge.writingconfig.driver.attack.count}{22}
\definevar{challenge.writingconfig.resources.codesnippets.count}{33}
\definevar{challenge.writingconfig.resources.docs.count}{27}

\definevar{challenge.discussoptions.count}{73}
\definevar{challenge.discussoptions.domain.os.count}{28}
\definevar{challenge.discussoptions.domain.serverapp.count}{31}
\definevar{challenge.discussoptions.securityaspect.deployment.count}{34}
\definevar{challenge.discussoptions.securityaspect.networking.count}{18}
\definevar{challenge.discussoptions.securityaspect.accesscontrol.count}{29}
\definevar{challenge.discussoptions.driver.attack.count}{25}
\definevar{challenge.discussoptions.driver.configuration.count}{22}
\definevar{challenge.discussoptions.resources.codesnippets.count}{23}
\definevar{challenge.discussoptions.resources.docs.count}{29}

\definevar{challenge.searching.count}{46}

\definevar{challenge.hardeningmeasure.count}{43}

\definevar{challenge.misconception.count}{19}

\definevar{challenge.faultyassumption.count}{17}

\definevar{domain.count}{7}
\definevar{domain.count.word}{seven}

\definevar{domain.serverapp.count}{92}
\definevar{domain.application.webserver.count}{45}
\definevar{domain.application.fileserver.count}{5}
\definevar{domain.application.emailserver.count}{5}

\definevar{domain.cloud.count}{24}
\definevar{domain.virtualization.count}{22}
\definevar{domain.os.count}{137}
\definevar{domain.supplychain.count}{16}
\definevar{domain.networkhardware.count}{8}
\definevar{domain.clientapp.count}{22}

\definevar{securityaspect.count}{7}
\definevar{securityaspect.count.word}{seven}

\definevar{securityaspect.deployment.count}{91}
\definevar{securityaspect.deployment.professional.count}{31}

\definevar{securityaspect.accesscontrol.count}{88}

\definevar{securityaspect.guides.count}{60}

\definevar{securityaspect.networking.count}{44}

\definevar{securityaspect.tooling.count}{32}

\definevar{securityaspect.malware.count}{10}

\definevar{securityaspect.crypto.count}{9}

\definevar{driver.count}{7}
\definevar{driver.count.word}{seven}
\definevar{driver.nodriver.count}{50}

\definevar{driver.attacks.count}{72}
\definevar{driver.attacks.vulnscan.count}{14}
\definevar{driver.attacks.tampering.count}{4}
\definevar{driver.attacks.spamming.count}{3}
\definevar{driver.attacks.incidents.count}{18}

\definevar{driver.external.count}{53}
\definevar{driver.external.compliance.count}{24}
\definevar{driver.external.distrustgov.count}{2}
\definevar{driver.external.pressure.count}{1}

\definevar{driver.config.count}{40}
\definevar{driver.config.reduceattacksurface.count}{5}
\definevar{driver.config.leastprivilege.count}{6}
\definevar{driver.config.defaultsettings.count}{2}
\definevar{driver.config.sandboxing.count}{3}

\definevar{driver.privacy.count}{27}
\definevar{driver.privacy.protecthideinformation.count}{7}
\definevar{driver.privacy.storesensitivedata.count}{2}
\definevar{driver.privacy.anonymity.count}{1}

\definevar{driver.automation.count}{16}

\definevar{driver.updatemigration.count}{13}

\definevar{date.collection}{June 2023}
\definevar{date.collection.end}{June 13, 2023}
\definevar{date.collection.start}{June 13, 2017}
\definevar{date.oldest_question}{September 30, 2017}
\definevar{date.newest_question}{December 13, 2022}
\definevar{date.years.collected}{6}
\definevar{date.years.collected.word}{six}

\newcommand{\postssecurity}{105}
\definevar{duplicates.postssecurity}{109} %
\newcommand{\postsstackoverflow}{85}
\newcommand{\postsserverfault}{41}
\newcommand{\postsunix}{31}
\definevar{duplicates.postsunix}{32} %
\newcommand{\postsaskubuntu}{9}
\definevar{duplicates.postsaskubuntu}{10} %
\newcommand{\postssitecore}{6}
\newcommand{\postssuperuser}{4}
\newcommand{\postsapple}{4}
\newcommand{\postsraspberrypi}{3}
\newcommand{\postscrypto}{3}
\newcommand{\postsdba}{2}
\newcommand{\postsdevops}{2}
\newcommand{\postsnetworkengineering}{2}
\newcommand{\postsdrupal}{2}
\newcommand{\postsjoomla}{2}
\definevar{duplicates.postsjoomla}{3} %
\newcommand{\postswordpress}{1}
\newcommand{\postsmonero}{1}
\newcommand{\poststor}{1}
\newcommand{\postsethereum}{1}
\newcommand{\postswebmasters}{1}
\newcommand{\postssoftwarerecs}{1}
\newcommand{\postscodegolf}{1}

\usepackage[acronym, %
    nohypertypes={acronym}, %
]{glossaries}

\newacronym{irb}{IRB}{Institutional Review Board}
\newacronym{erb}{ERB}{Ethical Review Board}
\newacronym{gdpr}{GDPR}{General Data Protection Regulation}
\newacronym{ips}{IPS}{Intrusion Prevention Systems}
\newacronym{ids}{IDS}{Intrusion Detections Systems}
\newacronym{aws}{AWS}{Amazon Web Services}

\newacronym{so}{SO}{Stack Overflow}
\newacronym{se}{SE}{Stack Exchange}
\newacronym{azure}{Azure}{Microsoft Azure}
\newacronym{gcp}{GCP}{Google Cloud Platform}

\newacronym{nlp}{NLP}{Natural Language Processing}

\newacronym{cis}{CIS}{Center for Internet Security}
\newacronym{stig}{STIG}{Security Technical Implementation Guides}
\newacronym{pcidss}{PCI-DSS}{Payment Card Industry Data Security Standard}

\newacronym{qda}{QDA}{qualitative data analysis}
\newacronym{irr}{IRR}{inter-rater reliability}

\newacronym{nist}{NIST}{U.S.\ National Institute of Standards and Technology}
\newacronym{bsi}{BSI}{German Federal Office for Information Security}

\begin{document}

\title{\Large \bf From Paranoia to Compliance:\\The Bumpy Road of System Hardening Practices on Stack Exchange}

\newcommand{\orcidSascha}{\orcidlink{0000-0002-5644-3316}}
\newcommand{\orcidAlex}{\orcidlink{0000-0003-2993-2568}}
\newcommand{\orcidDominik}{\orcidlink{TBD}}
\newcommand{\orcidHarjot}{\orcidlink{TBD}}
\newcommand{\orcidJan}{\orcidlink{0000-0002-6994-7206}}
\newcommand{\orcidJuliane}{\orcidlink{TBD}}
\newcommand{\orcidNico}{\orcidlink{TBD}}
\newcommand{\orcidNiklas}{\orcidlink{0000-0001-5621-8461}}
\newcommand{\orcidNoah}{\orcidlink{TBD}}
\newcommand{\orcidPhilip}{\orcidlink{0000-0002-8829-6074}}
\newcommand{\orcidSabrina}{\orcidlink{TBD}}
\newcommand{\orcidSandra}{\orcidlink{TBD}}
\newcommand{\orcidYas}{\orcidlink{0000-0001-7167-7383}}
\newcommand{\orcidJacques}{\orcidlink{0009-0007-8595-3706}}

\newcommand{\emailSascha}{sascha.fahl@cispa.de}
\newcommand{\emailAlex}{alexander.krause@cispa.de}
\newcommand{\emailDominik}{dominik.wermke@cispa.de}
\newcommand{\emailHarjot}{kaur@sec.uni-hannover.de}
\newcommand{\emailJan}{klemmer@sec.uni-hannover.de}
\newcommand{\emailJuliane}{juliane.schmueser@cispa.de}
\newcommand{\emailNico}{nicolas.huaman@cispa.de}
\newcommand{\emailNiklas}{niklas.busch@cispa.de}
\newcommand{\emailNoah}{noah.woehler@cispa.de}
\newcommand{\emailPhilip}{philip.klostermeyer@cispa.de}
\newcommand{\emailSabrina}{sabrina.amft@cispa.de}
\newcommand{\emailYas}{yasemin.acar@uni-paderborn.de}
\newcommand{\emailCISPA}{\{firstname.lastname\}@cispa.de}

\newcommand{\affiliationCISPA}{CISPA Helmholtz Center for Information Security, Germany}
\newcommand{\affiliationPAD}{Paderborn University, Germany}

\author{
    \IEEEauthorblockN{
        Niklas Busch\,\orcidNiklas{}\IEEEauthorrefmark{1},
        Philip Klostermeyer\,\orcidPhilip{}\IEEEauthorrefmark{1},
        Jan H.\ Klemmer\,\orcidJan{}\IEEEauthorrefmark{1},
        Yasemin Acar\,\orcidYas{}\IEEEauthorrefmark{2},
        Sascha Fahl\,\orcidSascha{}\IEEEauthorrefmark{1}
        }\\
        \IEEEauthorblockA{\IEEEauthorrefmark{1}\affiliationCISPA, \texttt{\emailCISPA}}\\
        \IEEEauthorblockA{\IEEEauthorrefmark{2}\affiliationPAD, \texttt{\emailYas}}
}

\maketitle

\begin{abstract}
    Hardening computer systems against cyberattacks is crucial for security.
However, past incidents illustrated, that many system operators struggle with effective system hardening.
Hence, many computer systems and applications remain insecure.
So far, the research community lacks an in-depth understanding of system operators' motivation, practices, and challenges around system hardening.
With a focus on practices and challenges, we qualitatively analyzed \var{posts.started_with_duplicates} \gls{se} posts related to system hardening.
We find that access control and deployment-related issues are the most challenging, and system operators suffer from misconceptions and unrealistic expectations.
Most frequently, posts focused on operating systems and server applications.
System operators were driven by the fear of their systems getting attacked or by compliance reasons.
Finally, we discuss our research questions, make recommendations for future system hardening, and illustrate the implications of our work.

\end{abstract}

\section{Introduction}\label{sec:intro}

System hardening is the process of deploying and configuring computer systems to keep adversaries out by identifying and addressing vulnerabilities~\cite{intelhardening}.
A study from~\citeauthor{rose_system_2020} illustrates that implementing system hardening benchmarks can effectively lower the risk of security breaches~\cite{rose_system_2020}.
Although a critical part of the job of system operators, system hardening is challenging and many struggle to deploy effective hardening measures~\cite{dietrich2018investigating, Mai_https_2022, Krombholz_Idea_2017}.
Hence, companies are regularly affected by successful cyberattacks facilitated by ineffectively hardened computer systems~\cite{MicrosoftBlog, cywarekozmolog, rapid7, lastpassincident, microsoft_breach_2022}.

In 2022, Microsoft experienced a data breach via unauthenticated access to a misconfigured Azure blob storage. 
As a result, sensitive company data from over 65,000~companies and 548,000~users was compromised~\cite{microsoft_breach_2022}.
Microsoft faced another severe data breach in 2023 when an attacker successfully stole a signing key transferred from the isolated production environment to the company's Internet-facing network due to another misconfiguration~\cite{MicrosoftBlog}.
\gls{aws} customers also face challenges regarding misconfigured services. 
In 2021, Cosmolog Kozmetik suffered a data breach that exposed the information of approximately 567,000 users due to incorrect configuration of~\gls{aws} S3 buckets~\cite{cywarekozmolog}. 
Rapid7's \emph{Cloud Misconfiguration Report 2022} revealed that attacks targeting services hosted on~\gls{aws} are particularly frequent, furthermore highlighting the need for increased vigilance in securely configuring and updating cloud-based infrastructure to protect sensitive data from unauthorized access~\cite{rapid7}.
Moreover, in 2022, LastPass suffered a breach caused by missed software updates and inadequate hardening processes~\cite{lastpassincident}.
Incidents like these emphasize the criticality and challenges around effective system hardening for improved computer security.
Previous research explored system hardening from various perspectives, as we elaborate in~\autoref{sec:relwork}. 
However, research lacks an in-depth understanding of system hardening-related practices, challenges, and guidance, as our findings suggest especially the latter to be troublesome. 
To better understand system operators' practices and challenges and provide actionable recommendations, we performed an in-depth analysis of~\var{posts.started_with_duplicates}~\gls{se} posts while consulting their answers for context to answer the following RQs:

\begin{description}
    \item[RQ1] \textit{What are common system hardening areas?} 
    System operators need to harden many different systems, services, and applications.
    We aim to identify the most common domains and security aspects in system hardening.
    \item[RQ2] \textit{What are drivers for system hardening?} 
    Motivations for system hardening are manifold. 
    We aim to understand better what drives system operators to strengthen their system and application security.
    \item[RQ3] \textit{What are challenges in system hardening?} 
    System or application misconfigurations can have severe security consequences.
    We are interested in identifying the most pressing challenges for system hardening.
\end{description}

To the best of our knowledge, our study is the first to investigate the challenges of hardening systems based on insights from the~\gls{se} platforms.
Using qualitative analysis, we examine 316~\gls{se} posts and identify six main domains and seven security aspects discussed in the field of system hardening, as well as six drivers that led to a post.
We pinpoint seven significant challenges that occur while system hardening and correlate them with domains, security aspects, and drivers.
Based on our findings, we make suggestions to address the identified challenges.
Further, we provide the dataset including the entire qualitative coding in our replication package\footnote{\url{https://doi.org/10.17605/OSF.IO/5DXAB}}.

\section{Related Work}\label{sec:relwork}

We discuss related work in three key areas: (i) security research using~\gls{se} data, (ii) system hardening with a technical focus, and (iii) human centered research in system hardening.
We also contextualize our contributions and highlight the novelty of our work.

\boldparagraph{Security Research with SE Data}

\gls{se} data, such as~\gls{so} posts, have been used profusely in past research, including research on security, privacy, and accompanying challenges. 
Most of the existing research strongly focuses on developers and programming languages.
Thus, we only deal with those related to system operators.

Properly implementing access control is essential to securing and hardening computer systems.
\citeauthor{xu_how_2017} examined system administrators' challenges in resolving access denial issues by examining 486 randomly selected posts mainly from~\gls{se} sites~\cite{xu_how_2017}.
They found frequent occurrences of security misconfigurations, where admins grant excessive access during troubleshooting, leading to potential security risks~\cite{xu_how_2017}.

Penetration testing is a method of verifying that system hardening measures were implemented correctly.
In 2019,~\citeauthor{rahman_birds_2019} conducted an empirical study to identify the knowledge needs of practitioners related to penetration testing by analyzing 548 questions from Information Security~\gls{se}~\cite{rahman_birds_2019}.
They identified five key knowledge needs, including a starting point, best practices, and legal concerns~\cite{rahman_birds_2019}.

However, the existing research only addressed sub-areas of system hardening. 
We argue that a comprehensive overview of the entire field of system hardening is needed to identify the main challenges in system hardening.

\boldparagraph{Technical System Hardening}

System hardening was mainly studied with a technical focus.
The operating system is one of the most important areas in system hardening.
Thus, much research was done on how to improve the operating system's security with hardening techniques.
Mostly, these papers suggested technical solutions like kernel hardening~\cite{abubakar_shard_2021, lin_grebe_2022}, custom kernel modules~\cite{tian_lbm_2019,wright_linux_2002} or improve current implementations with security features like the intel software guard extensions (SGX)~\cite{burihabwa_sgx-fs_2018, wang_running_2019,ozga_chors_2022}.

However, many papers also address a more practical approach to system hardening.
Containers became increasingly popular in recent years. 
Hence, much research has been done into hardening container environments.
\citeauthor{amith_raj_mp_enhancing_2016} propose hardening Docker deployments using virtualization techniques, automated testing, deployment tools, and security configuration management~\cite{amith_raj_mp_enhancing_2016}.
\citeauthor{dahlstrom_hardening_2019} suggest a lightweight container system approach to harden cross-domain applications~\cite{dahlstrom_hardening_2019}.
In addition to research on container hardening in particular, the possibility of adapting existing hardening measures from other areas was also investigated~\cite{amith_raj_mp_enhancing_2016, sun_security_2018, rose_system_2020}.

Guidelines and benchmarks strongly impact system hardening. 
Implementing hardening guidelines or standards is sometimes even a legal requirement for compliance in some environments.
From an incident perspective,~\citeauthor{rose_system_2020} presented five case studies involving large-scale security breaches to identify specific system hardening benchmarks that may have mitigated or even prevented the various attacks~\cite{rose_system_2020}.
\citeauthor{sasidharan_case_2022} presented another case study on the implementation of Windows system hardening using CIS controls, security tools, a security framework, and a remediation toolkit~\cite{sasidharan_case_2022}.
Regarding Ubuntu as a major distribution,~\citeauthor{sedano_auditing_2021} developed a tool to perform security audits based on 232~CIS benchmarks~\cite{sedano_auditing_2021}.
A newer study from 2024 by~\citeauthor{h_system_2024} showed that organizations encounter significant challenges while implementing the~\gls{cis} benchmark due to the complexity and time-consuming nature of manual configurations. 
They propose an efficient and scalable automation solution to streamline compliance, reduce errors, and enhance security~\cite{h_system_2024}.

This study confirms our finding that system hardening benchmarks are complex.
However, they don't work out which parts of system hardening are the most challenging and how guidelines and benchmarks are related to other obstacles and challenges in system hardening.
Furthermore, none of the papers picture the challenges of system hardening as a whole.

\boldparagraph{Human Centered Research in System Hardening}

In contrast to developers, system operators have received little attention in human centered research.
\citeauthor{dietrich2018investigating} conducted a mixed-methods study with system operators on misconfigurations. 
They showed that many participants were confronted with security misconfigurations, and one-third suffered from resulting incidents~\cite{dietrich2018investigating}.
A similar paper by~\citeauthor{Li_2019_keepers} examined the specific software update management practices of system administrators. 
They identified challenges and limitations in the update process and suggested future research directions~\cite{Li_2019_keepers}.
In an experiment~\citeauthor{Krombholz_Idea_2017} studied the challenges system operators face when deploying HTTPS certificates, revealing significant usability challenges~\cite{Krombholz_Idea_2017}.

While these studies, similar to our research, focus on the human factor, they mostly use interviews and surveys.
In contrast, our work analyzes reported challenges of experienced system operators and extends the current body of knowledge. 
For example, we demonstrate that system operators struggle to understand hardening measures.
The implementation of hardening measures is often only driven by regulations, such as guidelines and benchmarks. 
These guidelines are difficult to understand, resulting in operators blindly implementing measures without understanding the impact, leading to errors and misconceptions.

\section{Methodology}\label{sec:methodology}

\begin{figure}[ht]
    \centering
    \begin{tikzpicture}[>={Latex[width=1.5mm,length=1.5mm]},
    node distance = 0.3cm and 1.0cm,
    block/.style = {rectangle, rounded corners, draw=black, minimum width=8.0cm, minimum height=1cm, align=left, text width=80mm, inner sep=0.5em},
    recchannel/.style = {minimum width=1.1cm},
    auto,
    font=\scriptsize\sffamily
]

\node (collection) [block] {
\textbf{1.~Data Collection (\autoref{ssec:method-data-collection})} \\
Queries on data.stackexchange.com for posts on all communities: Having the tag \emph{hardening} -or- having \emph{hardening} -or- \emph{harden} in their title; resulting in \var{posts.allever} posts, no time constraint.
};

\node (filtering) [block, below =of collection] {
\textbf{2.~Filtering and Cleaning (\autoref{ssec:method-filtering-cleaning})} \\
Filtering and cleaning the dataset based on the community in which a post was created, and a temporal constraint of \var{date.years.collected.word} years backwards, resulting in 316 posts between \var{date.collection.start} and \var{date.collection.end}.
};

\node (initialcodebook) [block, below =of filtering] {
\textbf{3.1~Codebook Development: Initial Codebook (\autoref{sssec:method-initial-codebook})} \\
Two researchers agreed on an initial codebook based on the RQs and with consultation of related work giving best practices for naming of top-level code categories.};

\node (stabilizingcodebook) [block, below =of initialcodebook] {
\textbf{3.2~Codebook Development: Refining the Codebook (\autoref{sssec:method-codebook-refinement})}\\
Randomly selecting 98 posts; two researchers coded all posts, resolved conflicts and merged their codebooks for a first stable version of major categories and a vertical slice of insights.};

\node (coding) [block, below =of stabilizingcodebook] {
\textbf{4.~Iterative Coding (\autoref{ssec:method-coding})}\\
Three researchers coded all remaining 218 posts in teams of two; Iterative resolution of conflicts and minor revisions of the codebook if necessary.};

\node (mapping) [block, below =of coding] {
\textbf{5.~Affinity Diagramming (\autoref{ssec:method-affinity-diagramming})}\\
Using Affinity Diagramming for each major category of the codebook to group codes under similar topics, streamline the code set, and identify major categories to report in the analysis.};

\draw[->, very thick] (collection) -- (filtering);
\draw[->, very thick] (filtering) -- (initialcodebook);
\draw[->, very thick] (initialcodebook) -- (stabilizingcodebook);
\draw[->, very thick] (stabilizingcodebook) -- (coding);
\draw[->, very thick] (coding) -- (mapping);
\end{tikzpicture}
    \caption{Methodology of the study.}
    \label{fig:method}
\end{figure}
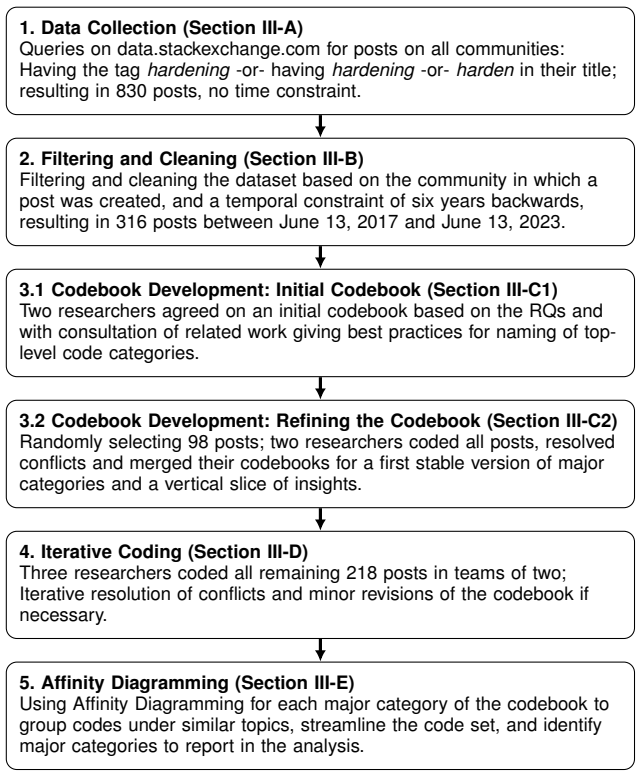

This section describes the data collection and qualitative analysis of~\var{posts.started_with_duplicates}~system hardening~\gls{se} posts. 
We describe the development of our codebook, the coding process, and the affinity diagramming procedure to identify groups and clusters based on relationships or similarities to support our analysis and the reporting of results. 
\autoref{fig:method} summarizes our procedure. 

\subsection{Data Collection}
\label{ssec:method-data-collection}

We used~\gls{se}'s API \emph{Data Explorer}~\cite{StackexchangeAPI} to collect~\gls{se} system hardening-related posts.

(1)~We collected posts with the tag \emph{hardening}. 
Tags~\cite{StackexchangeTags} are used on~\gls{se} to assign topics to a post and can be specified by the post creator. 
Alternatively, other users can suggest tags if they have the required reputation level. 
As tags are optional, not every post necessarily has a tag. 
On the \emph{Information Security}~\gls{se}~\cite{SecurityStackExchange} site, the \emph{hardening} tag is described as \enquote{the process of tightening security on a system}~\cite{SecurityStackExchangeHardeningTag}, which is in line with both the definition of system hardening (cf.\ \autoref{sec:intro}) and our research questions.

(2)~To compensate for untagged posts, we also included all posts containing the word \emph{hardening} or \emph{harden} in their title.
We searched for both, as the search terms \emph{harden} and \emph{hardening} resulted in different posts since the~\gls{se} API searches only for complete words but not sub-strings. %

We carefully chose the present filtering criteria. 
The chosen filtering has proven to be a good compromise between not missing many posts and keeping the false positive rate low.
First, we tried to use all posts containing the keyword \emph{hardening} or \emph{harden} anywhere in the post, not just the title and tags. 
This approach resulted in $\approx 2,200$ posts. 
A random analysis of the posts showed that many posts were off-topic. %
We also evaluated the tags used beside the \emph{hardening} tag in the posts in the dataset. 
We analyzed them by frequency and created bigrams and trigrams of tags to find relevant posts not tagged with hardening. 
However, we skipped this approach since the bigrams and trigrams contained overly generic tags such as \emph{apache}, \emph{operating system}, or \emph{security}.

Our approach is common practice in related research~\cite{tahaei_understanding_2020, imtiaz_challenges_2019, tahaei_privacy_2022, tahaei_understanding_2022, lopez_anatomy_2019}. %
Overall, we collected~\var{posts.allever}~posts.

\subsection{Filtering and Cleaning}
\label{ssec:method-filtering-cleaning}

Two researchers analyzed all posts by using their titles. 
All posts that seemed off-topic by their title were inspected in detail.
This revealed the need for additional filtering, and thus, we applied the following exclusion criteria.

(1)~Some posts were from non-technical~\gls{se} sites, like \emph{The Great Outdoors~\gls{se}}~\cite{OutdoorsStackExchange}. 
Upon scrutinizing these posts, it became evident that they did not align with the topic of our study. %
Consequently, we opted to narrow down our search only to technical~\gls{se} sites.

(2)~The search yielded some ancient posts, the oldest dating back to May 7 2009.
In examining some older posts, we noted the recurring problem of outdated information or treatment of outdated issues, such as the hardening of particularly outdated operating systems and versions. 
Therefore, and in line with similar research~\cite{tahaei_understanding_2022}, we excluded posts older than six years to examine current challenges appropriately.

\begin{figure}[t]
    \centering
    \includegraphics[width=\linewidth]{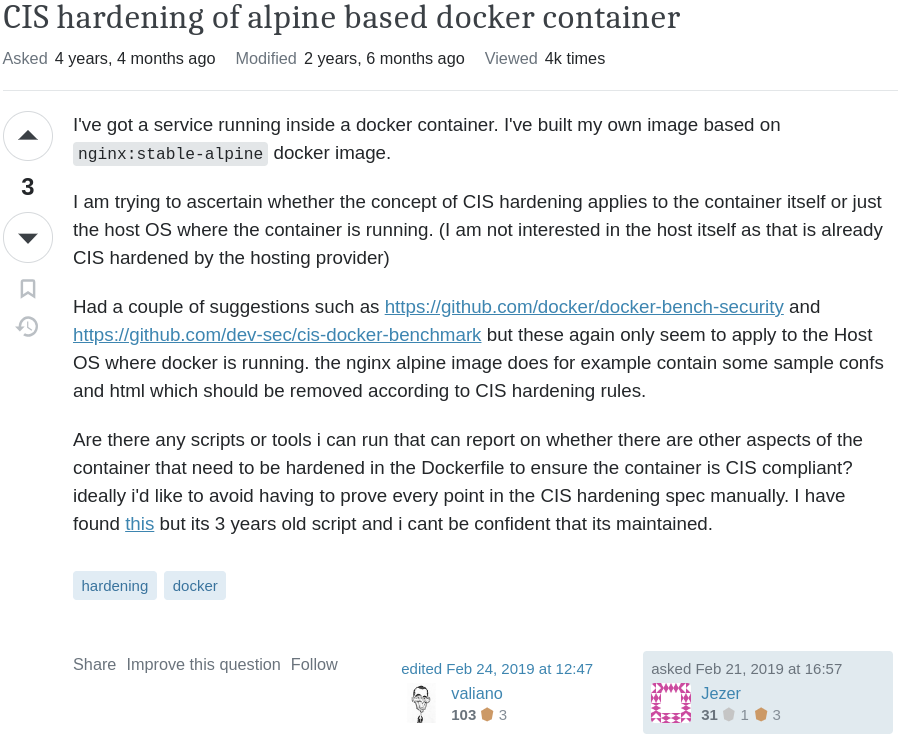}
\caption{Example post \sequestcite{204026}{https://security.stackexchange.com/questions/204026}{2019}{CIS hardening of alpine based docker container} from Information Security~\gls{se}.}
    \label{fig:example_post}
\end{figure}

At the time of collection in~\var{date.collection} and after filtering and cleaning, the final dataset included $N =\var{posts.started_with_duplicates}$ posts.
\autoref{fig:example_post} shows an example post from our dataset.

\autoref{tab:qeuscnt} lists the number of posts and their respective~\gls{se} site.

\subsection{Codebook Development}
\label{ssec:method-codebook-development}

In this section, we present the design of our initial codebook based on our research questions and its refinement by coding a random subset of posts from the dataset. %

\subsubsection{Initial Codebook} 
\label{sssec:method-initial-codebook}
We followed a semi-open coding approach.
Two researchers created an initial codebook based on the RQs and with lessons learned from related work, especially regarding top-level codebook categories~\cite{beyer_automatically_2018, tahaei_understanding_2020, andre_developers_2022}. 
Since our work aims to depict system hardening challenges system operators discuss online, we included 
(1)~the topic of the post, separated into general \emph{Domains}, such as \emph{Server Application} and security-related \emph{Security Aspects}, such as \emph{Access Control}, 
(2)~the specific \emph{Driver}, if any, in which a questioner expresses why they are asking their question, and 
(3)~the actual \emph{Challenge} a questioner faces. 
Furthermore, we included additional code categories, such as \emph{Context of Usage}, \emph{Technologies}, and \emph{Resources Provided}. 
These categories aimed to capture whether the post pertains to a professional or private environment, the specific technology involved, and any supplementary resources (\eg{} links, CLI commands, or error messages provided by the questioner), respectively. %

\subsubsection{Refining the Codebook} 
\label{sssec:method-codebook-refinement}
With the initial codebook, two researchers coded~\var{questions.sampled.sample}~randomly selected posts in an iterative semi-open coding approach~\cite{corbin1990grounded, strauss1997grounded, charmaz2014constructing}. 
We coded our observations by mutual agreement, adding, removing, and discussing subcodes, periodically resolving any conflicts, and verifying that the codebook correctly reflected the targeted elucidation of the RQs.

Two of these posts were marked as a \emph{Duplicate} of others on~\gls{se}.
In such cases,~\gls{se} links to the similar but older original post~\cite{StackexchangeDuplicates}.
Upon closer inspection, we decided to include both duplicate and associated older posts in the dataset. 
We reasoned that if a question was reposted during our analyzed period, the content of that question remained relevant.
Furthermore, we marked~\var{questions.unrelated.sample}~posts as \emph{Unrelated} when they were not related to hardening in the sense of our work. %

\subsection{Iterative Coding}
\label{ssec:method-coding}
Subsequently, three researchers continued coding the remaining~\var{questions.remaining} of the~\var{posts.started_with_duplicates}~posts in alternating teams of two.
They regularly merged their codebooks, discussed any changes, and resolved conflicts until reaching saturation~\cite{Birks2011, Urquhart2013}. 
We refrain from reporting an intercoder agreement because each conflict was resolved when it occurred~\cite{mcdonald2019reliability}. 
With~\var{questions.count.duplicates}~additions from posts marked as duplicate, our final dataset encompasses~\var{posts.started_with_duplicates}~posts.

\subsection{Affinity Diagramming}
\label{ssec:method-affinity-diagramming}

To finally sort our codebook and identify the most critical topics, we used \emph{affinity diagramming}~\cite{Beyer:1997:affinity} after completing the coding process. 
Affinity diagramming is a visual technique that identifies related topics of words or codes from a coding process and sorts them into common categories. 
Similar to our approach, \citeauthor{beyer2014}~\cite{beyer2014} used this method to find common topics in a manual categorization of Android application development issues on \gls{so}.
Using affinity diagramming, we identified six domains, seven security aspects, six drivers, and seven challenge categories. 

\subsection{Limitations}

We only collected posts from Q\&A sites in the~\gls{se} network for the study. 
We did not consider other Q\&A sites, forums, or communities where questions about system hardening might be discussed, such as Reddit or Twitter, and we note that since~\gls{se} provides a public forum where anyone can sign up and ask questions, participants in discussions may be rather uninitiated in the topic.
However, we argue that this is an acceptable trade-off, as~\gls{se} forms the primary discussion platforms for developers and administrators with large communities. 
Moreover, the posts on Q\&A sites from the~\gls{se} network have been used as a data source by researchers of various disciplines in the past and have generated many relevant and valuable insights in the past~\cite{beyer2014,Fischer:2017,zhang_are_2018,ahasanuzzaman_classifying_2018,tahaei_understanding_2020,tahaei_privacy_2022,tahaei_understanding_2022,meng_secure_2018}. 

Our dataset only contains~\gls{se} posts tagged with \emph{hardening} or having \emph{hardening} or \emph{harden} in their title. 
Hence, we could have missed posts that were related in content with hardening but were not tagged or named accordingly. 

Our analysis only encompasses questions asked in~\gls{se} posts. 
Assuming that system operators only ask questions on~\gls{se} when they seek and cannot find a solution for some challenge, we only have insights on these publicly discussed challenges.
However, we might miss challenges that are not discussed,~\eg{} questions that seem too trivial and were therefore not asked at all, or those that system operators could solve themselves without creating a post.

\subsection{Ethical Considerations}

Since all the analyzed data is publicly available, and therefore it is no human subjects research, this work did not require \gls{irb} approval. 
\gls{se} encourages researchers to use its data to produce academic papers~\cite{SOPapers}, licenses posts under \emph{Creative Commons ShareAlike}~\cite{Stackexchangecreativecommons} and requires researchers to attribute posts with a direct link~\cite{SOAttribution} (as we do throughout the paper, also for the reader's convenience, cf.\ \autoref{tab:stackexchange}).

\section{Results}\label{sec:results}

This section presents our results, starting with a dataset description and discussed domains.
We furthermore report on discussed security aspects (\autoref{subsec:secaspects}), drivers for asking questions (\autoref{subsec:drivers}), and the system hardening challenges we identified (\autoref{subsec:challenges}).
Since a post on~\gls{se} may contain several (sub-)questions, the total number of coded observations does not equal the total number of posts in the dataset. 

While we report numbers throughout this paper, we note that this is exploratory qualitative research and the numbers should therefore not be interpreted as quantitative statistical results. 
Instead, they are intended to give an impression of the weight of a theme.
We report an absolute count and the relative percentage over the whole dataset for each domain, security aspect, and challenge.

\boldparagraph{About the Dataset} 
Our final dataset includes~\var{posts.started_with_duplicates}~posts from between~\var{date.collection.start}, and~\var{date.collection.end}. 
As of~\var{date.collection.end}~\var{posts.answered} of the posts were answered from which~\var{posts.answered.accepted} had an accepted answer~\cite{SecurityStackExchangeAcceptedAnswer}. 
\var{posts.unanswered} posts were unanswered, excluding comments. 
We found~\var{posts.closed} \emph{closed} posts.
Most commonly, the post was closed because the question was already answered in another post (\var{questions.count.duplicates}).
We included them in our analysis and eventually added their linked post to the dataset. 
Other reasons for closing the posts were \emph{needs more focus} (\var{posts.needsmorefocus}), \emph{not suitable for this site} (\var{posts.notsuitable}), \emph{needs details or clarity} (\var{posts.clarity.and.opinionbased}), and \emph{opinion-based} (\var{posts.clarity.and.opinionbased}).

\boldparagraph{Context of Usage} 
We have identified 65 posts with a professional context and only 17 with a private one. 
We could not assign the context for 139~posts.

\boldparagraph{Domains}\label{subsec:domains}

\begin{table*}[tbp]
        \centering
        \scriptsize
        \caption{Domains and their occurrence in the dataset.}
        \label{tab:domains}
        \renewcommand{\arraystretch}{1.33}
        \setlength{\tabcolsep}{0.6\tabcolsep}
        \setlength{\defaultaddspace}{0.1\defaultaddspace} %
        \rowcolors{2}{white}{gray!10}
        \begin{tabularx}{\linewidth}{lXrl}
            \toprule
            \textbf{Domain} & \textbf{Description} & \multicolumn{2}{l}{\textbf{Observations}} \\
            \midrule
            \textbf{Operating System} & Posts on configuring or administering operating systems and services, including Linux, Windows, macOS, and Android & \var{domain.os.count} &(\printpercentoverall{\var{domain.os.count}}) \\
            \textbf{Server Application} & Posts about server applications such as Web-, Mail-, and File-servers & \var{domain.serverapp.count} &(\printpercentoverall{\var{domain.serverapp.count}})\\
            \textbf{Cloud} & Posts on managing and configuring cloud providers like \gls{aws}, \gls{azure}, and \gls{gcp} & \var{domain.cloud.count} &(\printpercentoverall{\var{domain.cloud.count}}) \\
            \textbf{Client Application} & Posts about client and workstation applications such as browsers and Office & \var{domain.clientapp.count} &(\printpercentoverall{\var{domain.clientapp.count}})\\
            \textbf{Virtualization} & Posts on virtualized technologies such as virtual machines and containers & \var{domain.virtualization.count} &(\printpercentoverall{\var{domain.virtualization.count}})\\
            \textbf{Supply Chain} & Supply chain related posts like package management and configuration management (\eg~Ansible) & \var{domain.supplychain.count} &(\printpercentoverall{\var{domain.supplychain.count}})\\
            \bottomrule
        \end{tabularx}
\end{table*}

System hardening topics were discussed broadly across various domains, ranging from operating systems and server applications over client applications, cloud-related posts, posts regarding virtualization, and posts about the supply chain. 
\autoref{tab:domains} accompanies each domain with its corresponding description and the frequency count indicating how often it has been observed.

\subsection{Security Aspects}\label{subsec:secaspects}

The security aspect indicates to which security sub-area a post belongs and helps to understand which security aspects are challenging in system hardening. 
Posts covered the~\var{securityaspect.count.word} security aspects \emph{Deployment}, \emph{Access Control}, \emph{Guidelines and Benchmarks}, \emph{Networking}, \emph{Security Tooling}, \emph{Malware} and \emph{Cryptography}. 
Furthermore, eight posts were left without a security aspect assigned. 
This occurred due to the broad nature of these posts. %

\boldparagraph{Deployment (\printstats{\var{securityaspect.deployment.count}})}
This category contains posts dealing with deploying applications, services, systems, and specific configurations. 
Comprehensive deployment care is critical for system hardening,~\eg{} to prevent the creation of attack vectors through insecure initial configurations.
We found that operating systems were the most common domain (38), followed by server applications (35).
Web servers were discussed especially within server applications (23).
Deploying systems in the cloud (\printstatsc{13}{91}) or on virtualization hypervisors (\printstatsc{12}{91}) also kept users busy. 
Another range of posts on virtualization deployment dealt with container systems such as Docker or their orchestration (9).
The software supply chain, especially package management (8), was another recent topic.
Over one-third of the posts came up in a professional context (\printstatsc{\var{securityaspect.deployment.professional.count}}{\var{securityaspect.deployment.count}}).

\boldparagraph{Access Control (\printstats{\var{securityaspect.accesscontrol.count}})}
One-fourth of the posts concerned access control challenges. %
The majority were in the domain of operating systems (51). %
Permissions were the most common access control issue within operating systems (27), and Linux was most frequently mentioned as the technology used as an operating system (39).
Access control was a significant security aspect within the domain of client applications (11).
Browsers were the most mentioned client applications related to access control.

\boldparagraph{Guidelines and Benchmarks (\printstats{\var{securityaspect.guides.count}})}
The importance of system hardening guidelines and benchmarks,~\eg{} due to regulatory requirements, and related challenges for system operators are also reflected in our data.
Those serve two primary purposes: first, to ensure compliance with a standard such as the~\gls{pcidss}~\cite{pcidss}; and second, to outline best practices for hardening specific systems or services, as exemplified by the~\gls{cis} benchmark.
The~\gls{cis} benchmarks describe itself as \blockquote[\cite{cis}]{prescriptive configuration recommendations \textelp{that} represent the consensus-based effort of cybersecurity experts}. 
We found 32 posts that address CIS benchmarks.
These guidelines and benchmarks primarily pertain to various operating systems and applications, representing their main focus domains. 
Operating systems accounted for 29 posts of these posts, while server applications constituted 22 posts.

\boldparagraph{Networking (\printstats{\var{securityaspect.networking.count}})}
The security aspect of networking contains all posts dealing with communication between systems or services over a network, including regulation and firewall filtering. 
The predominant domain in networking was server applications, with 20 occurrences.
In contrast to the other security aspects instead of web servers, the posts mainly addressed SSH servers (\printstatsc{8}{\var{securityaspect.networking.count}}), file servers (\printstatsc{5}{\var{securityaspect.networking.count}}) and mail servers (\printstatsc{3}{\var{securityaspect.networking.count}}). 
Firewalling comprised 19 of the posts.
Furthermore, 11 posts were asked about firewalling on Linux systems in particular.
The mainly used firewall under Linux with 7 occurrences was iptables~\cite{netfilter} or its successor nftables~\cite{netfilter}.
For Windows, three posts related to firewalls.

\boldparagraph{Security Tooling (\printstats{\var{securityaspect.tooling.count}})}
Posts dealt with tools such as~\gls{ids},~\gls{ips}, audit tools, or monitoring tools.
The domains of security tools are very diverse, and we could not identify particular trends. 
The majority of posts focused on audit (\printstatsc{8}{\var{securityaspect.tooling.count}}) or monitoring tools and logging tools (\printstatsc{7}{\var{securityaspect.tooling.count}}).
These tools were related to guidelines and benchmarks (\printstatsc{11}{\var{securityaspect.tooling.count}}). %

\boldparagraph{Malware (\printstats{\var{securityaspect.malware.count}})}
Malware played a relatively minor role in our dataset.
These posts primarily focused on the potential risks of malware infections, with some addressing remedial measures for compromised systems.
Web servers (\printstatsc{6}{\var{securityaspect.malware.count}}) were the predominant domain.
Three of these posts highlighted instances where WordPress was utilized as the entry point for malware intrusions into the system.
One focused on malware operating mechanisms, and another on whether malware concerns Linux users.

\boldparagraph{Cryptography (\printstats{\var{securityaspect.crypto.count}})}
Few posts were cryptography related.
Cryptographic cipher suites were the topic in nearly all posts (\printstatsc{7}{\var{securityaspect.crypto.count}}),
ranging from choosing the best cipher suites over removing deprecated ones to how to configure them.
Questions focused on server applications, such SSH servers, web servers, and mail servers.
These posts included code snippets from configuration files (6) or links to external websites or documents (6).

\begin{summaryBox}{Summary: Security Aspects}
    \ We identified deployment and access control as essential areas for system hardening.
    Guidelines emerged as essential, whereas networking, security tooling, malware, and cryptography were less critical.
\end{summaryBox}

\subsection{Drivers}
\label{subsec:drivers}

Overall, we found~\var{driver.count.word}~different drivers that motivated users to seek help on~\gls{se}. 
Drivers included fear of attacks, external factors, configuration purposes, privacy, automation, and updates and migrations. 
We could not identify drivers for~\var{driver.nodriver.count}~posts. %
System hardening is always done to enhance the protection of a system or service and thus to prevent an attack.
Hence, we focused on coding explicitly mentioned drivers.

\boldparagraph{Fear of Attacks (\printstats{\var{driver.attacks.count}})}

The fear of attacks was the most common driver in our dataset. 
Past security incidents (\printstatsc{18}{\var{driver.attacks.count}}) were most common. %
Users sought assistance and guidance to address vulnerabilities and remediate past attacks and mitigate the risk of future attacks:  
\blockquote[231046]{My WP site just got hacked for the third time even after following WP hardening guidelines \textelp{} How can I prevent future attacks?}
Additionally, system operators often mentioned security and audit scans (\printstatsc{14}{\var{driver.attacks.count}}) as they desire to identify and address potential vulnerabilities proactively.
\blockquote[56143561]{We have an Apache \textelp{} in production. The security Audit team found few vulnerabilities lately which needs to be fixed.}
In related posts, system operators asked to protect their systems against vulnerability scans.
\blockquote[54459236]{I am trying to change the default path of the WP default directories such as \emph{wp-content, wp-include} etc to avoid \emph{wpscan}}
System operators were also driven to ask questions by various minor types of attacks, including data tampering, privilege escalations, and side-channel attacks.

\boldparagraph{External (\printstats{\var{driver.external.count}})}
External drivers emerged as the second most prominent motivation. 
Users were motivated by various external reasons to enhance the security of their systems. 
Compliance with regulations and adherence to guidelines and standards was a central reason for the extrinsically motivated implementation of security measures (\printstatsc{46}{\var{driver.external.count}}). 
\blockquote[941192]{In our company, we want to configure our Windows-based infrastructure compliant to the IASE SCAP specifications, e.g., the Microsoft Windows Server 2016 STIG Benchmark.} or
\blockquote[204026]{\textelp{} are \textins{there} other aspects of the container that need to be hardened in the Dockerfile to ensure the container is~\gls{cis} compliant?}

We identified further external drivers.
Distrust in government served as a driver for system operators, prompting them to enhance their system security proactively, \eg to be safe against intelligence agencies. 
\blockquote[191469]{What would be the best practices for securing a single-purpose Windows laptop against a determined foreign intelligence agency from tampering with data on the machine?}
Other external drivers were user pressure, reading about system hardening best practices, or attending security courses or lectures.

\boldparagraph{Configuration (\printstats{\var{driver.config.count}})}

Configuration revolved \ie around implementing security principles, such as least privilege (\printstatsc{16}{\var{driver.config.count}}), or sandboxing and isolation (\printstatsc{6}{\var{driver.config.count}}).
For applying the least privilege principle, users asked to restrict user privileges, as in \blockquote[222616]{\textelp{} my requirement is that these root users not have access to data which is located under certain directories.} or limit the permissions of processes, as in \blockquote[1090794]{The idea is to block everything and allow only what is actually known to be used by the server services.}.

Other posts focused on reducing the attack surface of a system. 
Therefore, users sought to decrease risk by removing unused software and services or disabling unnecessary features (\printstatsc{11}{\var{driver.config.count}}). 
\blockquote[969212]{how to \textelp{} disable Apache when only using Tomcat?} or \blockquote[219981]{look at attack surface, and remove/deny/disable ereything that an attacker could use to escape the jail or pivot to other networked devices.}

Additionally, (\printstatsc{6}{\var{driver.config.count}}) posts discussed inherent distrust in default settings. 
System operators were driven by a distrust of default settings that could leave their systems vulnerable to attacks, \ie
\blockquote[83606]{Microsoft isn't shipping Windows Defender with the strongest settings.}

\boldparagraph{Privacy (\printstats{\var{driver.privacy.count}})}
The privacy driver primarily revolved around protecting and hiding sensitive information (\printstatsc{19}{\var{driver.privacy.count}}). 
System operators focused on safeguarding different types of data.
This was mainly data that an attacker could use to gather details about the target system or network, like the version details of used software.
\blockquote[182508]{I would like to modify Win OS banner to defeat OS detection from scanning tools like \emph{nmap} for example.} or, \blockquote[195359]{
My organization wants to restrict all the plugins/tools like Netcraft and Builtwith to detect all the server-side technologies for security reasons like platform, operating system name and version, web server name and version}
They also prioritized securing documents and application data, such as API interfaces or logs. 
\blockquote[406843]{
I want to make it as difficult as possible to extract information from my hard drive if stolen or lost as there are sensitive documents and details on it.} 
or
\blockquote[215398]{I'd prefer to enable as much logging as possible but secure the access to the logs.}
By implementing privacy-enhancing measures, system operators aimed to ensure the confidentiality and integrity of sensitive information.

\boldparagraph{Automation (\printstats{\var{driver.automation.count}})}

The automation driver stems from the desire to automate the hardening process to save time and effort.
Furthermore, automation allows for consistent and efficient deployment of security configurations across multiple systems, reducing the risk of human error and ensuring adherence to security standards. 
Many system operators seek to automate the implementation of security measures by searching or developing custom hardening scripts.
\blockquote[955208]{I'm trying to write a hardening script to remove the cron directory in an alpine Linux based docker image.}
Another approach for automatization is utilizing configuration management tools like Ansible \blockquote[69471721]{I want to be able to modify specific local policies on my WS 2019. I've tried to use the win\_security\_policy module from ansible \textelp{}}.
These automation efforts often rely on guidelines and benchmarks as their foundation.
\blockquote[68112625]{Anyone has any Ansible or other scripts to perform~\gls{cis} hardening level on the above spec?}

\boldparagraph{Update and Migration (\printstats{\var{driver.updatemigration.count}})}

After migrations, changes in the system environment can render existing hardening instructions ineffective or outdated, requiring system operators to adapt and modify their hardening measures accordingly. 
\blockquote[73808977]{I'm working on a hardening task of RHEL 8. The step now is set umask Daemon, I've tried to find /etc/sysconfig.init file to add umask 027 but it's not exist likes RHEL 7. Where can I config this umask on RHEL 8?}
Similarly, system updates aimed at improving security can sometimes result in unexpected issues or conflicts that impact the functionality of specific components.
\blockquote[645077]{\textelp{} everything has been working fine up until a while ago after an update \textelp{}}

\begin{summaryBox}{Summary: Drivers}
    \ Attack prevention or remediation was the most common driver in our dataset.
    The second most pressing driver was compliance,~\eg{} to hardening standards.
\end{summaryBox}

\subsection{Challenges in System Hardening}
\label{subsec:challenges}

This section presents the categories for characterizing the seven system hardening challenges we identified.
We connect them accordingly, as those can be related to challenges, domains, security aspects, and drivers. %
Our findings range from curiosity about needing explanations at a higher level to asking for assistance with implementing or executing a concrete task and troubleshooting. 

\boldparagraph{Question Triggers in System Hardening}
\begin{figure*}[htb]
    \centering
    \includegraphics[width=0.8\textwidth]{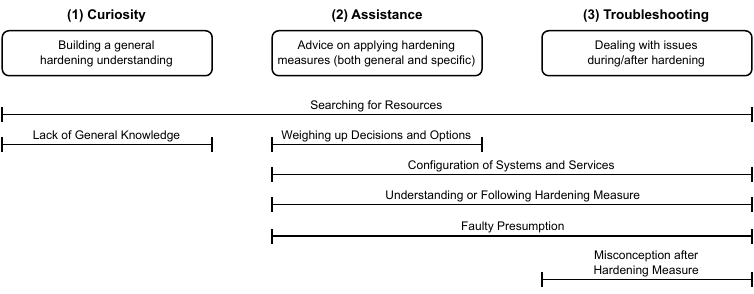}
    \caption{An overview on question triggers and associated hardening challenges.}
    \label{fig:stages_of_system_hardening}
\end{figure*}

We identified three triggers for asking questions on system hardening, namely (1)~\emph{Curiosity}, (2)~\emph{Assistance}, and (3)~\emph{Troubleshooting}, as shown in \autoref{fig:stages_of_system_hardening}.
This is likely due to the overall hardening process, which starts with general questions that come up while searching for information on hardening measures, followed by more concrete questions about implementing those measures, and ends with highly concrete questions about encountered errors and problems.
These triggers come with an overall decreasing level of abstraction,~\ie{} troubleshooting questions are typically very concrete compared to more general questions triggered by curiosity.
We also added the identified challenges to \autoref{fig:stages_of_system_hardening} to indicate which stage a challenge typically occurs. 

The first trigger, \emph{Curiosity}, describes questions that come up in a discovery phase: users were curious about a topic and tended to ask higher-level, theoretical, and abstract questions. 
These users' challenges can be described as a lack of general knowledge or searching for resources to gather mostly initial information. 
The second trigger, \emph{Assistance}, refers to questions relating more concretely to a specific topic or software. 
Questioners usually already have a vague to concrete idea of a hardening measure they want to apply but ask a related question because they need help with it.
The challenges associated with this trigger include discussing and weighing options, writing configuration files, and executing the proper steps to apply hardening measures.
Hence, it covers all questions concerning the implementation of hardening measures.
The third and last trigger, \emph{Troubleshooting}, consists of questions where a hardening action may have failed.
For example, system operators might get stuck when their hardening measure does not work as expected due to an error or unexpected event.

\boldparagraph{Co-Occurences}

\begin{figure*}[htb]
     \centering
     \begin{subfigure}[t]{0.47\linewidth}
         \centering
         \includegraphics[width=0.9\linewidth]{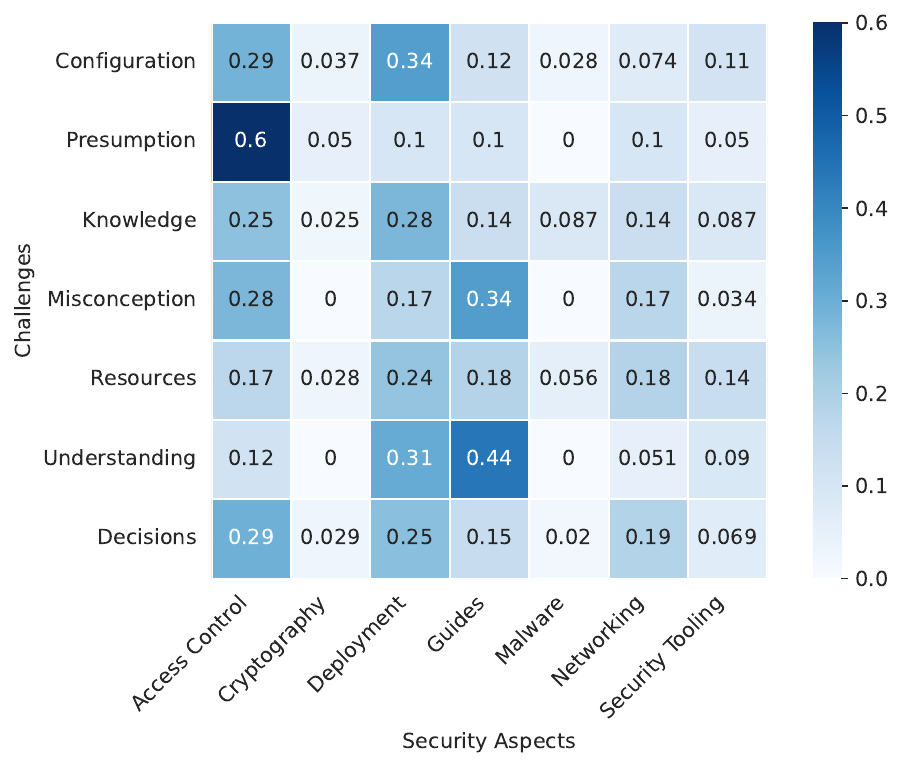}
        \caption{Row-wise normalized co-occurrence of challenges and security aspects.}
        \label{fig:heatmap_challenges_secaspects}
     \end{subfigure}
     \hfill
     \begin{subfigure}[t]{0.47\linewidth}
         \centering
         \includegraphics[width=0.9\linewidth]{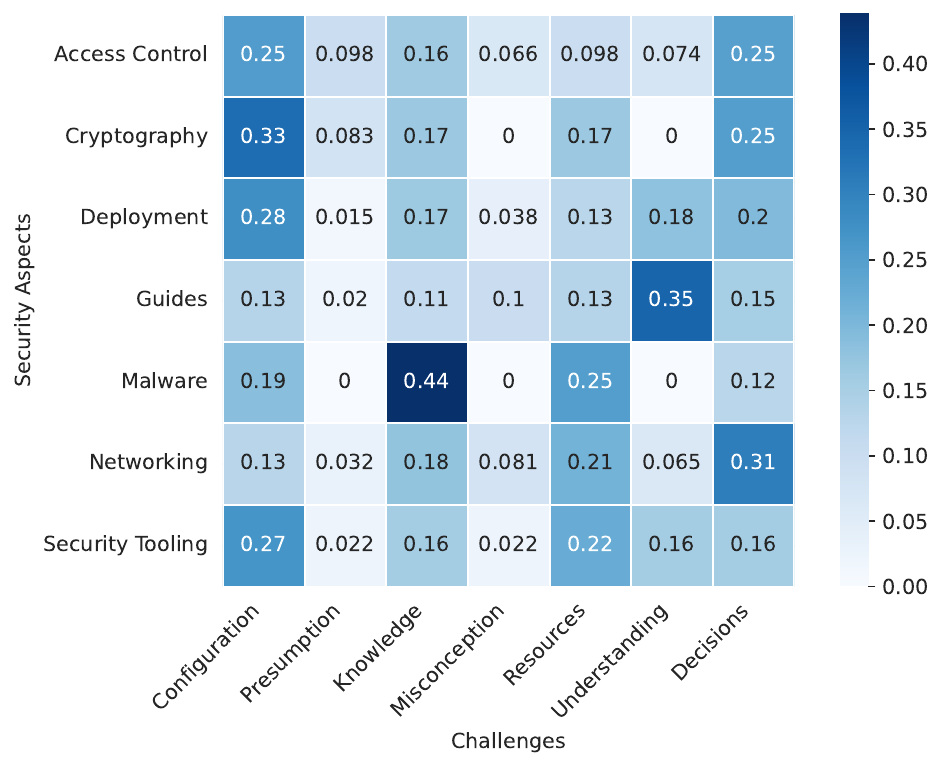}
         \caption{Row-wise normalized co-occurrence of security aspects and challenges.}
         \label{fig:heatmap_secaspects_challenges}
     \end{subfigure}
         \caption{Visualization of the relative frequencies and associations between Security Aspects and Challenges and vice versa using row-wise normalization, to give insights into their co-occurrence dynamics within the dataset. Challenge abbreviations: 
         \textit{Configuration:} Configuration of Systems and Services, 
         \textit{Presumption:} Faulty Presumption, 
         \textit{Knowledge:} Lack of General Knowledge, 
         \textit{Misconception:} Misconception after Hardening Measure, 
         \textit{Resources:} Searching for Resources, 
         \textit{Understanding:} Understanding or Following Hardening Measures, 
         \textit{Decisions:} Weighing up Decisions and Options, 
         \textit{Guides:} Guidelines and Benchmarks.}
    \label{fig:challenges_heatmaps}
\end{figure*}

Below, we illustrate the relationships between challenges, domains, and security aspects in our dataset.
\autoref{fig:heatmap_challenges_secaspects} illustrates the normalized co-occurrence of security aspects within the challenges, while \autoref{fig:heatmap_secaspects_challenges} illustrates the normalized co-occurrence of challenges within the security aspects.
The figures illustrate the relative frequencies at which specific security aspects occur in each challenge. 
By normalizing the data over the domains, the figure allows a comparative analysis of the prevalence and distribution of security aspects, highlighting each aspect's relative importance and prominence within the different domains.
Below, we report the figure's insights in detail.

\begin{figure}[htb]
    \includegraphics[width=0.9\linewidth]{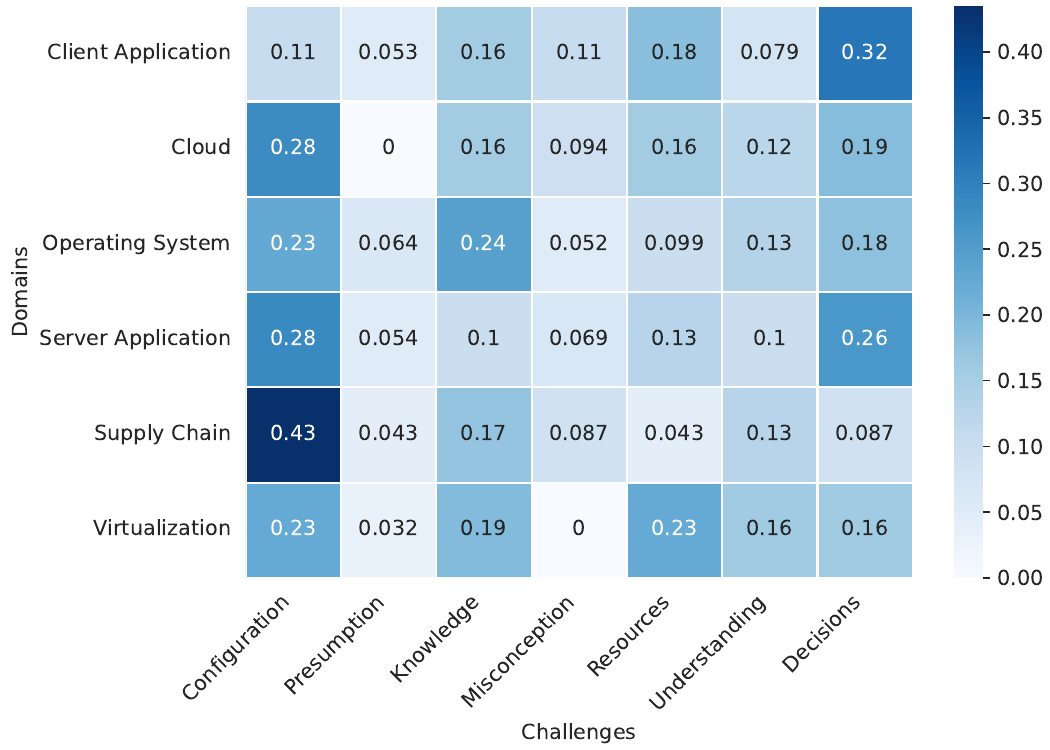}
    \caption{Row-wise normalized co-occurrence of Domains and Challenges}
    \label{fig:heatmap_domains_challenges}
\end{figure}

\boldparagraph{Searching for Resources (\printstats{\var{challenge.searching.count}})}

The challenge of searching for resources is unique in that it can occur due to all three triggers of system hardening (cf.\ \autoref{fig:stages_of_system_hardening}). 
This challenge is about finding relevant and reliable resources on system hardening, which is crucial before, during, and after the implementation of hardening measures. 
These resources encompass both abstract elements, such as formal guidelines and best practices, and concrete components, like system images, tools, and specialized software. 
The posts often seek resources on specific topics,~\eg{} Windows Defender: \blockquote[83606]{\textins{Is} someone knowing a Tool/Software, which is recommended for hardening Windows Defender. \textins{sic!}}

\autoref{fig:heatmap_challenges_secaspects} indicates that the security aspect of \emph{Deployment} was most commonly associated with this challenge (\printstatsc{17}{\var{challenge.searching.count}}).
These posts were predominantly related to the \emph{Virtualization} domain, as depicted in \autoref{fig:heatmap_domains_challenges},~\eg{} for searching Docker images, like \blockquote[195845]{Is there any service that provides certified, security-hardened Docker images for common platforms like Python, PHP, Node, Java, etc. with 0 major/critical CVEs. \textins{sic!}}.

System operators also faced challenges when searching for resources concerning \emph{Guidelines and Benchmarks} (\printstatsc{13}{\var{challenge.searching.count}}) in combination with security tooling. 
They typically sought tools to implement security measures outlined in a guide or to perform system audits to comply with guidelines. 
Examples of such posts include:
\blockquote[75371459]{Is there any way to attain a bash script that would allow me to automate the installation of~\gls{cis} security policy to existing Oracle Linux 8?}
or
\blockquote[70463445]{I’m trying to find any open source tool or scripts available for direct use to audit the Windows 2019 system against the~\gls{cis} benchmarks \textelp{}}.
Therefore, the driver \emph{Being Compliant to some Standard} (\printstatsc{11}{\var{challenge.searching.count}}) was a major reason to search for resources.

\boldparagraph{Lack of General Knowledge (\printstats{\var{challenge.knowledge.count}})}

Some questions seemed to be triggered by a curiosity about system hardening (\autoref{fig:stages_of_system_hardening}), and cover building a general understanding of system hardening.
More than half of the posts were about the domain \textit{Operating System}, mostly covering \textit{Linux/Unix}.
Therefore, many posts were about kernel features and default Unix commands. 
However, components of the operating system, such as the file system, also came up frequently.  
Therefore, specific knowledge of operating systems seems to be a significant challenge for hardening computer systems. 
\textit{Deployment} is the \textit{Security Aspect} about which the most questions were asked (cf.\ \autoref{fig:heatmap_challenges_secaspects}), closely followed by \textit{Access Control}. 
This also aligns with the predominant domain \textit{Operating System}, as many questions addressed specific configuration details.

\blockquote[210589]{I am hardening CentOS/RHEL 7.6. The hardening documents recommend disabling the automounter, "unless it is necessary." Why is autofs such a problem? One of the benefits of networking is a shared file system. What other alternatives are there?}

In addition, this challenge also contains most posts without an identifiable \textit{Security Aspect} or \textit{Driver}. 

\boldparagraph{Weighing up Decisions and Options (\printstats{\var{challenge.discussoptions.count}})}

The challenge of \textit{Weighing up decisions and options} occurs by questions that were triggered due to requesting \textit{Assistance} in system hardening (\autoref{fig:stages_of_system_hardening}). 
Users often expressed an understanding of their goal but required assistance to make specific decisions or choose between multiple options.

This challenge covers diverse domains. 
Most posts covered access control.
The most significant uncertainty in decisions was about permissions.

\blockquote[57731428]{We are running a Spring Boot application that we start up with a simple \emph{'java -jar jarFile'}, and the image is built using maven's dockerfile-maven-plugin. With that being said, should I be changing the user to an unprivileged user before running that[...]?}

Furthermore, networking, especially firewalls, appeared frequently (\autoref{fig:heatmap_challenges_secaspects}). 
Here alternative firewalls were sought, or the question arose whether additional firewall rules would increase security: 
\blockquote[487876]{I wonder if it is useful also to set the policies to DROP for mangle, raw, and security tables (not nat table because it does not work) to more secure the server?}
These findings align with the driver \emph{Fear of Attacks} mentioned in 25~posts. In 14 cases, this fear of attacks resulted from previous security incidents.

\boldparagraph{Configuration of Systems and Services (\printstats{\var{challenge.writingconfig.count}})}

This challenge concerns writing or finding configurations for hardening operating systems or network components and services.
Therefore, this challenge can be triggered by the need for \textit{Assistance} when the questioner asks for help before or during implementation.
Moreover, this challenge also came up in \textit{Troubleshooting} posts, when errors occurred,~\eg{} caused by an incorrect configuration.

The domain with the most challenges in this category was \emph{Server Applications} (\autoref{fig:heatmap_domains_challenges}). 
Web servers were by far the most common. 
\emph{Deployment} was the most common security aspect covered in 34~posts (\autoref{fig:heatmap_challenges_secaspects}):
\blockquote[969212]{How would I completely disable Apache, since we aren't using it at all? If I do the above would it impact tomcat in any way? (I'm assuming not). How would I alternatively keep apache httpd running and just redirect all requests to tomcat? What files should I put these redirect rules in? httpd/conf.d/redirect.conf \textelp{}???}

The fear of attacks often drove questioners. 
Being attacked, such as following automated network scans for vulnerabilities, was of particular concern to the questioners: 
\blockquote[54459236]{I am trying to change the default path of the WP default directories such as wp-content, wp-include etc to avoid wpscan. I have tried using plugin would it possible to perform the same using manual techniques. I am using apache as a web server.
An example, I have tried: \\
\emph{RewriteRule \textasciicircum cms\_plugins/(.+) /wordpress/wp-content/plugins/\$1 [L,QSA]}}
Another driver was the configuration of services to reduce the attack surface or implementing the least privilege principle.

\boldparagraph{Understanding or Following Hardening Measure (\printstats{\var{challenge.hardeningmeasure.count}})}

This challenge was triggered by posts related to both \textit{Assistance} and \textit{Troubleshooting}, wherein questioners encounter difficulty in comprehending specific hardening measures despite having prior knowledge of their intended implementation. 
Typically, queries about operating systems or server applications were addressed in this.

This challenge is noteworthy due to the highest prevalence of security guidelines and benchmarks in numerous posts and the inclusion of external references or links in 35 of these posts (\autoref{fig:heatmap_challenges_secaspects}). 
Most frequently, posts mentioned the~\gls{cis} benchmark, which often resulted in errors requiring further clarification and assistance: 
\blockquote[1018828]{I'm hardening fedora OS following the~\gls{cis} benchmark for fedora 28. In one of the remediations, the Benchmark provides a script that modifies the files \emph{system-auth} and \emph{password-auth}. 
When I apply the changes with \emph{authselect apply-changes} I get an error because the files were modified. 
Supposedly I can modify these files, but I'm not understanding how to commit the changes. 
I've been searching about this but am still stuck}

Furthermore, certain users acknowledged their need for more security proficiency. %
Users were mostly extrinsically motivated due to reaching compliance with a standard -- e.g. the~\gls{cis} benchmark:
\blockquote[899084]{First of all I would like to say I'm not a Linux/Solaris guy, but just assigned task to look at 1 particular item in the hardening checklist, so thinking of seeking help here to understand more.}

\boldparagraph{Faulty Presumption (\printstats{\var{challenge.faultyassumption.count}})}

This challenge contains posts triggered by some incorrect assumption. Users requested \textit{Assistance}, or \textit{Troubleshooting}. %

Thematically, these issues primarily addressed operating system challenges (11~posts). 
Access control was most frequent in 12~questions (\autoref{fig:heatmap_challenges_secaspects}).
Often users misunderstood certain technologies worked and how they could be implemented, like in this~\gls{se} post: \blockquote[1460813]{How to protect gnome-terminal or any shell with a password and maybe something like recaptcha... It could consult a shadowed password database or require the user to log in like tty}
Users, like the one above, were often driven by the fear of attacks.

In this challenge, many questioners (7) indicated a professional background. %

\boldparagraph{Misconception after Hardening Measure (\printstats{\var{challenge.misconception.count}})}

This challenge addresses issues where users do not understand what a hardening action has done and why errors occur. 
Posts were triggered by \textit{Troubleshooting} due to something unexpected occurring after the questioner tried to apply a hardening measure.

Most challenges occurred in the operating systems domain.
Windows has an above-average share of 33.3\%. 
The most important security aspect was guidelines and benchmarks with 10~questions (\autoref{fig:heatmap_domains_challenges}). 
This indicates that many guidelines are implemented blindly, without being aware of what individual actions do:
\blockquote[199246]{I am enforcing a hardening policy on my organization's workstations. One of the policies I removed \textelp{} is called "Allow system to be shut down without having to log on". 
Users started to complain and asked us to re-enable this policy, and I tend to agree. 
Can you think of a good reason why to disable?}

The quote shows that hardening recommendations for server systems have been applied to client systems.
Another security aspect that came up more often than average was \textit{Firewalling} under \textit{Networking}. 
Users implemented firewall rules without properly understanding their consequences and then tried to investigate root causes for errors, such as:
\blockquote[1027188]{As part of a "Hardening" task, I need to run\\
\emph{iptables -P INPUT DROP \\
iptables -P OUTPUT DROP \\
iptables -P FORWARD DROP \\}
on our servers. Normally we would run this command and then run to implement the new policy. However, as soon as I ran \emph{iptables -P  OUTPUT DROP} my SSH disconnected. Is this due to the OS being RHEL? How do I configure this machine to allow my IP address through?}

The most frequent driver was \emph{Being compliant to some standard}. 
This illustrates that system operators might not fully understand the guidelines and benchmarks such as~\gls{cis}, and consequently struggle to successfully apply hardening measures. 

\begin{summaryBox}{Summary: System Hardening Challenges}
    \ The two major challenges were configuring systems and services and weighing up hardening decisions and options.
    Overall, some questioners have clear hardening objectives but need help accomplishing them.
\end{summaryBox}

\section{Discussion}\label{sec:discussion}

Below, we discuss our findings and outline the potential implications of our work.

\boldparagraph{RQ1} \textit{What are common system hardening areas?} 
We identified certain trends toward specific topics among the system hardening questions on~\gls{se}. 
Overall, topics related to operating systems and server applications were often discussed while cloud operations, providers, and virtualization were less frequently discussed. 
Furthermore, many questions discussed the deployment of measures. 
In particular, access control and permissions were actively discussed, including user, application, and file system object permissions. 
The findings on existing guidelines and benchmarks were surprising, as they are widely used but often misleading.
Uncertainties about the use of specific security tools are very concerning to users.
This includes audits and the interpretation of the output of such audit tools, as well as questions from the networking area, particularly firewalls.

\boldparagraph{RQ2} \textit{What are drivers for system hardening?} 
Many questions revealed that system operators wanted to protect themselves from both abstract and specific attack patterns. 
While those can be preventive, others are entirely reactive:
Some~\gls{se} system hardening posts revealed that security incidents had occurred previously -- being the primary reason for seeking system hardening advice to prevent future incidents. 
Related to this were audits and security scans that imposed challenges to~\gls{se} users. 
In general, however, the biggest external driver was compliance with standards, guidelines, policies, or benchmarks. 
In 50 posts no driver could be determined.
For example because the questioners did not elaborate on their motives, and the questions were often very brief and focused on the questioner's problems without explaining the context, which admittedly is not always necessary.
Other interesting drivers came from special configurations to reduce the target's attack surface or implement the least-privilege principle. 
Smaller categories concerned data protection, especially the driver for protecting sensitive information, the automation of hardening tasks, and the occurrence of updates or migrations that presented system operators with new challenges and led to a preoccupation with hardening.

\boldparagraph{RQ3} \textit{What are the challenges in system hardening?} 
One of the most discussed system hardening challenges centers around configurations for systems and applications.
This challenge was often associated with access control issues and the deployment of configurations in operating systems and server applications, affecting both Linux and Windows. 
Other posts related to software supply chains,~\eg{} challenges securing package managers, or cryptography setups,~\eg{} searching best practices for configuring TLS cipher suites.

In addition, users discussed and evaluated certain hardening options on~\gls{se} when they could not assess which decision would be the right one in their case, \eg~ weighing up on which firewall should be used.
Interestingly, many of these discussions concerned client applications. 
Unsurprisingly, we identified challenges around official system hardening guidelines and benchmarks like~\gls{cis}. 
These difficulties are often related to following and applying measures from such guidelines. 
In particular, even small deviations from the system's behavior as described in these guidelines have led to problems.
We also noted general discussions about specific attacks, malware, and the desire to protect systems without having a specific realization of the hardening in mind.

The search for additional guidance and resources was fairly evenly distributed across various topics. 
This includes questions related to data privacy,~\ie{} best practices to prevent leaking sensitive information, such as from web servers.
Misjudgments after implementing a hardening measure were frequently related to hardening guides and benchmarks, with the CIS benchmark again to the fore. 
We also identified misconceptions about system hardening. 
Especially in access control and networking, system operators with misconceptions asked for help.
In these posts, they had to be made aware of their misconceptions and explain the errors resulting from them.

\subsection{Recommendations}
\label{ssec:recommendations}

\boldparagraph{Secure Default Settings}

Overall, our findings indicate a demand for improving security -- often because default security settings were considered secure enough. 
Hence, we recommend that applications be preconfigured with strong and safe defaults.
For instance, we found that the default cipher suites do not align with best practices for SSH servers and TLS deployment in web servers, like Apache~\cite {mozilla_ssl_generator}.
Consequently,~\gls{se} users asked for support to disable outdated and insecure TLS cipher suites or, even worse, might be completely unaware of this security problem.
Providing secure default settings could prevent these pitfalls.
If needed, these secure defaults can still be explicitly changed and downgraded,~\eg{} to ensure compatibility with legacy systems. 
Moreover, applications could ship with different configurations tailored to various security levels, like in the Mozilla SSL Configuration Generator~\cite{mozilla_ssl_generator}.
The advantage of secure defaults is that they are proactive and require little to no user interaction or security expertise.
A proactive and user-friendly approach to system hardening can be achieved by emphasizing secure defaults and providing customization options.

\boldparagraph{Documentation}
Our findings imply that the availability and accessibility of comprehensive and easy-to-use documentation pose significant challenges.
This aligns with prior research on documentation in secure software development~\cite{Acar:2016ww}.
Official documentation provided by the system or software provider lacks effective hardening advice. 
Hence, system operators struggle to find system hardening advice. %
In many cases, relying on external resources, such as the CIS benchmark~\cite{cis}, becomes necessary.
This reliance on external sources adds another layer of complexity to system hardening efforts, as they are often inaccessible or costly.
We also recommend improving the documentation by providing comprehensive, user-friendly guides with actionable recommendations to greatly support system operators in their pursuit of effective system hardening practices.

\boldparagraph{Security Operation Champions}

Similar to the role of a security champion in software development teams~\cite{Weir_needs_2020}, we recommend a dedicated security operation champion role in system operator teams to support system hardening. %
They could promote a proactive approach to system hardening and prioritize security considerations throughout the process. 
These dedicated individuals should possess in-depth knowledge of system vulnerabilities and hardening techniques, allowing them to guide and educate other system operators. 
This might enable organizations to establish a culture of continuous improvement in system hardening, effectively mitigating risks, and enhancing overall security posture.

\subsection{Implications}

Our results suggest that system operators invest significant efforts to harden their systems. 
While some may have no strategy, others possess some sort of guideline. 
Our findings indicate that system operators face difficulties locating or correctly implementing system hardening guidelines.
Specifically, posts on web servers concerning guides in a broader sense are also quite frequent, particularly those about the CIS benchmark. 
Given the popularity of these guidelines, it is not surprising that they are used extensively in system hardening. 
Nevertheless, the frequency of problems regarding guides suggests that comprehension and other challenges may be associated with these guidelines.
Issues range from the initial search for an appropriate guide to challenges that may arise later in the hardening process, such as the occurrence of non-functioning features after the application of hardening measures. 
Furthermore, in some posts, additional side effects of measures have been reported that are no longer covered by the guide. 
Finding relevant guides can be difficult, and the guides tend to be intended for a more advanced audience that is expected to possess a certain degree of literacy regarding using system commands for administration.
Many users have also expressed a desire to automate hardening measures.
However, we observed that this can result in complications, as some individuals may struggle to understand the consequences of the automated execution of typical hardening guidelines.

\section{Conclusion}\label{sec:conclusion}

We analyzed~\var{posts.started_with_duplicates} system hardening related Stack Exchange questions to gain insights into system operators' motivations, practices, and challenges. 
Our results show that misconfigurations are significant challenges, particularly in access control and provisioning. 
Furthermore, finding, understanding, and following established hardening measures leads to a higher incidence of questions indicating issues. 
Questions centered mostly around operating system and server application hardening. 
Motivations for system hardening questions include fear of attacks and compliance requirements. 
Our analysis recommends using secure default settings, improving system documentation, and user-centric approaches with security operation champions to enhance system security.

\printbibliography

@misc{intelhardening,
	title        = {{What Is System Hardening?}},
	author       = {{Intel Corporation}},
	note         = {Accessed: 2023-05-05},
	howpublished = {\url{https://www.intel.com/content/www/us/en/business/enterprise-computers/resources/system-hardening.html}}
}

@misc{mozilla_ssl_generator,
	title        = {{moz://a SSL Configuration Generator}},
	author       = {Mozilla Foundation},
	note         = {Accessed: 2023-05-05},
	howpublished = {\url{https://ssl-config.mozilla.org/}}
}

@misc{lastpassincident,
	title        = {{LastPass Hack: Engineer's Failure to Update Plex Software Led to Massive Data Breach}},
	author       = {{The Hacker News}},
	year         = 2023,
	note         = {Accessed: 2023-05-05},
	howpublished = {\url{https://thehackernews.com/2023/03/lastpass-hack-engineers-failure-to.html}}
}

@inproceedings{Weir_needs_2020,
	title        = {{From Needs to Actions to Secure Apps? The Effect of Requirements and Developer Practices on App Security}},
	author       = {Charles Weir and Ben Hermann and Sascha Fahl},
	year         = 2020,
	month        = 8,
	booktitle    = {29th USENIX Security Symposium (USENIX Security 20)},
	publisher    = {USENIX Association},
	pages        = {289--305},
	isbn         = {978-1-939133-17-5},
	url          = {https://www.usenix.org/conference/usenixsecurity20/presentation/weir}
}

@inproceedings{beyer2014,
	title        = {{A Manual Categorization of Android App Development Issues on Stack Overflow}},
	author       = {Beyer, Stefanie and Pinzger, Martin},
	year         = 2014,
	booktitle    = {2014 IEEE International Conference on Software Maintenance and Evolution},
	pages        = {531--535},
	doi          = {10.1109/ICSME.2014.88}
}

@book{Beyer:1997:affinity,
	title        = {{Contextual Design: Defining Customer-Centered Systems}},
	author       = {Beyer, Hugh and Holtzblatt, Karen},
	year         = 1997,
	publisher    = {Morgan Kaufmann Publishers Inc.},
	address      = {San Francisco, CA, USA},
	isbn         = 9780080503042
}

@misc{StackexchangeAPI,
	title        = {{Stack Exchange API}},
	author       = {{Stack Exchange}},
	year         = 2022,
	note         = {Accessed: 2023-05-05},
	howpublished = {\url{https://api.stackexchange.com/}}
}

@misc{SOAttribution,
	title        = {{Attribution Required}},
	author       = {Atwood, Jeff},
	year         = 2009,
	note         = {Accessed: 2023-05-05},
	howpublished = {\url{https://stackoverflow.blog/2009/06/25/attribution-required/}}
}

@misc{SOPapers,
	title        = {{Academic Papers Using Stack Overflow Data}},
	author       = {Atwood, Jeff},
	year         = 2009,
	note         = {Accessed: 2023-05-05},
	howpublished = {\url{https://stackoverflow.blog/2010/05/31/academic-papers-using-stack-overflow-data/}}
}

@misc{cywarekozmolog,
	title        = {{{Cosmolog Kozmetik} Leaks Half a Million Customers' Data Due to Cloud Misconfiguration}},
	author       = {{cyware}},
	year         = 2021,
	note         = {Accessed: 2023-05-05},
	howpublished = {\url{https://cyware.com/news/cosmolog-kozmetik-leaks-half-a-million-customers-data-due-to-cloud-misconfiguration-362e7416}}
}

@misc{rapid7,
	title        = {{2022 Cloud Misconfigurations Report}},
	author       = {{rapid7}},
	year         = 2022,
	note         = {Accessed: 2023-05-05},
	howpublished = {\url{https://www.rapid7.com/info/cloud-misconfigurations-research-report/}}
}

@misc{microsoft_breach_2022,
	title        = {{Misconfigured Azure Blob Storage Exposed the Data of 65K Companies and 548K Users}},
	author       = {{Sumeet Wadhwani}},
	year         = 2022,
	howpublished = {\url{https://www.spiceworks.com/it-security/cloud-security/news/microsoft-azure-cloud-misconfiguration/}}
}

@misc{MicrosoftBlog,
	title        = {{Results of Major Technical Investigations for Storm-0558 Key Acquisition}},
	author       = {{MSRC}},
	year         = 2023,
	note         = {Accessed: 2023-09-14},
	howpublished = {\url{https://msrc.microsoft.com/blog/2023/09/results-of-major-technical-investigations-for-storm-0558-key-acquisition/}}
}

@misc{SecurityStackExchange,
	title        = {{Information Security}},
	author       = {Stackexchange.com},
	note         = {Accessed: 2023-05-05},
	howpublished = {\url{https://security.stackexchange.com/}}
}

@misc{OutdoorsStackExchange,
	title        = {{The Great Outdoors}},
	author       = {Stackexchange.com},
	note         = {Accessed: 2023-05-05},
	howpublished = {\url{https://outdoors.stackexchange.com/}}
}

@misc{StackexchangeTags,
	title        = {{What are tags, and how should I use them?}},
	author       = {Stackexchange.com},
	note         = {Accessed: 2023-05-05},
	howpublished = {\url{https://meta.stackexchange.com/help/tagging}}
}

@misc{StackexchangeDuplicates,
	title        = {{Why are some questions marked as duplicate?}},
	author       = {Stackexchange.com},
	note         = {Accessed: 2023-05-05},
	howpublished = {\url{https://stackoverflow.com/help/duplicates}}
}

@misc{Stackexchangecreativecommons,
	title        = {{Terms of Service}},
	author       = {Stackexchange.com},
	note         = {Accessed: 2023-05-05},
	howpublished = {\url{https://stackoverflow.com/legal/terms-of-service/\#licensing}}
}

@misc{SecurityStackExchangeHardeningTag,
	title        = {{About hardening - Tag Info}},
	author       = {security.stackexchange.com},
	note         = {Accessed: 2023-05-05},
	howpublished = {\url{https://security.stackexchange.com/tags/hardening/info}}
}

@misc{SecurityStackExchangeAcceptedAnswer,
	title        = {{What does it mean when an answer is \"accepted\"?}},
	author       = {{Stackoverflow.com}},
	note         = {Accessed: 2023-05-05},
	howpublished = {\url{https://stackoverflow.com/help/accepted-answer}}
}

@misc{cis,
	title        = {{CIS Benchmarks List}},
	author       = {{Center for Internet Security (CIS)}},
	note         = {Accessed: 2023-05-05},
	howpublished = {\url{https://www.cisecurity.org/cis-benchmarks}}
}

@misc{pcidss,
	title        = {{System Hardening Standards for Complying with PCI DSS}},
	author       = {{PCI Security Standards Council (PCI SSC)}},
	note         = {Accessed: 2023-05-05},
	howpublished = {\url{https://www.pcidssguide.com/system-hardening-standards-for-complying-with-pci-dss/}}
}

@misc{netfilter,
	title        = {{Netfilter -- firewalling, NAT, and packet mangling for Linux}},
	author       = {netfilter.org},
	note         = {Accessed: 2023-05-05},
	howpublished = {\url{https://www.netfilter.org/}}
}

@book{Birks2011,
	title        = {{Grounded Theory: A Practical Guide}},
	author       = {Birks, Melanie and Mills, Jane},
	year         = 2011,
	month        = {01},
	isbn         = {978-1-84860-992-1}
}

@book{Urquhart2013,
	title        = {{Grounded Theory for Qualitative Research: A Practical Guide}},
	author       = {Urquhart, Cathy},
	year         = 2013,
	month        = {01},
	doi          = {10.4135/9781526402196},
	isbn         = 9781847870544
}

@inproceedings{Li_2019_keepers,
	title        = {{Keepers of the Machines: Examining How System Administrators Manage Software Updates For Multiple Machines}},
	author       = {Frank Li and Lisa Rogers and Arunesh Mathur and Nathan Malkin and Marshini Chetty},
	year         = 2019,
	month        = 8,
	booktitle    = {Fifteenth Symposium on Usable Privacy and Security (SOUPS 2019)},
	publisher    = {USENIX Association},
	address      = {Santa Clara, CA},
	pages        = {273--288},
	isbn         = {978-1-939133-05-2},
	url          = {https://www.usenix.org/conference/soups2019/presentation/li}
}

@inproceedings{Mai_https_2022,
	title        = {{Are HTTPS Configurations Still a Challenge?: Validating Theories of Administrators' Difficulties with TLS Configurations}},
	author       = {Mai, Alexandra and Schedler, Oliver and Weippl, Edgar and Krombholz, Katharina},
	year         = 2022,
	publisher    = {Springer-Verlag},
	address      = {Berlin, Heidelberg},
	pages        = {173–193},
	doi          = {10.1007/978-3-031-05563-8_12},
	isbn         = {978-3-031-05562-1},
	url          = {https://doi.org/10.1007/978-3-031-05563-8\%5F12},
	numpages     = 21
}

@inproceedings{Krombholz_Idea_2017,
	title        = {{"I Have No Idea What I{\textquoteright}m Doing" - On the Usability of Deploying {HTTPS}}},
	author       = {Katharina Krombholz and Wilfried Mayer and Martin Schmiedecker and Edgar Weippl},
	year         = 2017,
	month        = 8,
	booktitle    = {26th USENIX Security Symposium (USENIX Security 17)},
	publisher    = {USENIX Association},
	address      = {Vancouver, BC},
	pages        = {1339--1356},
	isbn         = {978-1-931971-40-9},
	url          = {https://www.usenix.org/conference/usenixsecurity17/technical-sessions/presentation/krombholz}
}

@String { acm              = {ACM} }

@String { ieee             = {IEEE} }

@String { nist             = {NIST} }

@String { sage             = {SAGE Publications} }

@String { taylor           = {Taylor and Francis} }

@String { usenix           = {USENIX Association} }

@String { ssp16            = {Proc.\ 37th IEEE Symposium on Security and Privacy (SP'16)} }

@String { ssp17            = {Proc.\ 38th IEEE Symposium on Security and Privacy (SP'17)} }

@String { cscw             = {Computer Supported Cooperative Work (CSCW)} }

@String { sp               = {IEEE Security \& Privacy } }

@String { zfs              = {{Zeitschrift f{\"u}r Soziologie}} }

@inproceedings{Acar:2016ww, 
    author                  = {Acar, Yasemin and Backes, Michael and Fahl, Sascha and Kim, Doowon and Mazurek, Michelle L and Stransky, Christian},
    title                   = {{You Get Where You're Looking For: The Impact of Information Sources on Code Security}},
    booktitle               = ssp16,
    publisher               = ieee,
    year                    = {2016},
}

@inproceedings{Fischer:2017, 
    author                  = {Fischer, Felix and B{\"o}ttinger, Konstantin and Xiao, Huang and Stransky, Christian and Acar, Yasemin and Backes, Michael and Fahl, Sascha},
    title                   = {{Stack Overflow Considered Harmful? The Impact of Copy\&Paste on Android Application Security}},
    booktitle               = ssp17,
    publisher               = ieee,
    year                    = {2017}
}

@article{corbin1990grounded,
    title                   = {{Grounded theory research: Procedures, canons and evaluative criteria}},
    author                  = {Corbin, Juliet and Strauss, Anselm},
    journal                 = zfs,
    volume                  = {19},
    number                  = {6},
    pages                   = {418--427},
    year                    = {1990}
}

@book{strauss1997grounded,
    title                   = {Grounded theory in practice},
    author                  = {Strauss, Anselm and Corbin, Juliet M},
    year                    = {1997},
    publisher               = sage
}

@book{charmaz2014constructing,
    title                   = {{Constructing Grounded Theory}},
    author                  = {Charmaz, Kathy},
    year                    = {2014},
    publisher               = sage
}

@misc{Nist, 
	title =		{Nist},
	url =		{https://www.nist.gov/},
}

@misc{java,
    title={{Java main page}},
    url={https://www.java.com/},
}

@misc{php,
    title={{PHP main page}},
    url={http://php.net/},
}

@article{mcdonald2019reliability,
  title={Reliability and inter-rater reliability in qualitative research: Norms and guidelines for CSCW and HCI practice},
  author={McDonald, Nora and Schoenebeck, Sarita and Forte, Andrea},
  journal={Proceedings of the ACM on Human-Computer Interaction},
  volume={3},
  number={CSCW},
  pages={1--23},
  year={2019},
  publisher={ACM New York, NY, USA}
}

@inproceedings{dietrich2018investigating,
  title={Investigating system operators' perspective on security misconfigurations},
  author={Dietrich, Constanze and Krombholz, Katharina and Borgolte, Kevin and Fiebig, Tobias},
  booktitle={Proceedings of the 2018 ACM SIGSAC Conference on Computer and Communications Security},
  pages={1272--1289},
  year={2018}
}

@inproceedings{sedano_auditing_2021,
	title        = {{Auditing {Linux} {Operating} {System} with {Center} for {Internet} {Security} ({CIS}) {Standard}}},
	author       = {Sedano, Wadlkur Kurniawan and Salman, Muhammad},
	year         = 2021,
	month        = jul,
	booktitle    = {2021 {International} {Conference} on {Information} {Technology} ({ICIT})},
	pages        = {466--471},
	doi          = {10.1109/ICIT52682.2021.9491663},
}

@inproceedings{amith_raj_mp_enhancing_2016,
	title        = {{Enhancing security of {Docker} using {Linux} hardening techniques}},
	author       = {{Amith Raj MP} and Kumar, Ashok and Pai, Sahithya J and Gopal, Ashika},
	year         = 2016,
	booktitle    = {2016 2nd {International} {Conference} on {Applied} and {Theoretical} {Computing} and {Communication} {Technology} ({iCATccT})},
	publisher    = {IEEE},
	address      = {Bangalore, India},
	pages        = {94--99},
	doi          = {10.1109/ICATCCT.2016.7911971},
	isbn         = {978-1-5090-2399-8},
	url          = {http://ieeexplore.ieee.org/document/7911971/},
	urldate      = {2022-12-06}
}

@inproceedings{dahlstrom_hardening_2019,
	title        = {{Hardening {Containers} for {Cross}-{Domain} {Applications}}},
	author       = {Dahlstrom, Jason and Brock, Jim and Tenaw, Mekedem and Shaver, Matthew and Taylor, Stephen},
	year         = 2019,
	month        = nov,
	booktitle    = {{MILCOM} 2019 - 2019 {IEEE} {Military} {Communications} {Conference} ({MILCOM})},
	publisher    = {IEEE},
	address      = {Norfolk, VA, USA},
	pages        = {1--6},
	doi          = {10.1109/MILCOM47813.2019.9020992},
	isbn         = {978-1-72814-280-7},
	url          = {https://ieeexplore.ieee.org/document/9020992/},
	urldate      = {2022-12-06}
}

@inproceedings{rose_system_2020,
	title        = {{System {Hardening} for {Infrastructure} as a {Service} ({IaaS})}},
	author       = {Rose, Tina and Zhou, Xiaobo},
	year         = 2020,
	month        = jul,
	booktitle    = {2020 {IEEE} {Systems} {Security} {Symposium} ({SSS})},
	publisher    = {IEEE},
	address      = {Crystal City, VA, USA},
	pages        = {1--7},
	doi          = {10.1109/SSS47320.2020.9174202},
	isbn         = {978-1-72814-316-3},
	url          = {https://ieeexplore.ieee.org/document/9174202/},
	urldate      = {2022-12-06},
}

@article{tahaei_understanding_2022,
	title        = {{Understanding {Privacy}-{Related} {Advice} on {Stack} {Overflow}}},
	author       = {Tahaei, Mohammad and Li, Tianshi and Vaniea, Kami},
	year         = 2022,
	journal      = {Proc. Priv. Enhancing Technol.},
	volume       = 2022,
	number       = 2,
	pages        = {114--131},
	doi          = {10.2478/popets-2022-0038},
}

@inproceedings{tahaei_privacy_2022,
	title        = {{Privacy, {Permissions}, and the {Health} {App} {Ecosystem}: {A} {Stack} {Overflow} {Exploration}}},
	shorttitle   = {Privacy, {Permissions}, and the {Health} {App} {Ecosystem}},
	author       = {Tahaei, Mohammad and Bernd, Julia and Rashid, Awais},
	year         = 2022,
	booktitle    = {{EuroUSEC} 2022: {European} {Symposium} on {Usable} {Security}, {Karlsruhe}, {Germany}, {September} 29 - 30, 2022},
	publisher    = {ACM},
	pages        = {117--130},
	doi          = {10.1145/3549015.3555669},
}

@article{andre_developers_2022,
	title        = {{Developers {Struggle} with {Authentication} in {Blazor} {WebAssembly}}},
	author       = {Andr\'{e}, Pascal Marc and Sti\'{e}venart, Quentin and Ghafari, Mohammad},
	year         = 2022,
	doi          = {10.48550/ARXIV.2208.00258},
	url          = {https://arxiv.org/abs/2208.00258},
	urldate      = {2022-12-16},
	copyright    = {Creative Commons Attribution 4.0 International},
	note         = {Publisher: arXiv Version Number: 1},
}

@inproceedings{beyer_automatically_2018,
	title        = {{Automatically classifying posts into question categories on stack overflow}},
	author       = {Beyer, Stefanie and Macho, Christian and Pinzger, Martin and Di Penta, Massimiliano},
	year         = 2018,
	month        = may,
	booktitle    = {Proceedings of the 26th {Conference} on {Program} {Comprehension}},
	publisher    = {ACM},
	address      = {Gothenburg Sweden},
	pages        = {211--221},
	doi          = {10.1145/3196321.3196333},
	isbn         = {978-1-4503-5714-2},
	url          = {https://dl.acm.org/doi/10.1145/3196321.3196333},
	urldate      = {2022-12-02},
	language     = {en},
}

@inproceedings{ahasanuzzaman_classifying_2018,
	title        = {{Classifying stack overflow posts on {API} issues}},
	author       = {Ahasanuzzaman, Md and Asaduzzaman, Muhammad and Roy, Chanchal K. and Schneider, Kevin A.},
	year         = 2018,
	month        = mar,
	booktitle    = {2018 {IEEE} 25th {International} {Conference} on {Software} {Analysis}, {Evolution} and {Reengineering} ({SANER})},
	pages        = {244--254},
	doi          = {10.1109/SANER.2018.8330213},
}

@inproceedings{lopez_anatomy_2019,
	title        = {{An {Anatomy} of {Security} {Conversations} in {Stack} {Overflow}}},
	author       = {Lopez, Tamara and Tun, Thein and Bandara, Arosha and Mark, Levine and Nuseibeh, Bashar and Sharp, Helen},
	year         = 2019,
	month        = may,
	booktitle    = {2019 {IEEE}/{ACM} 41st {International} {Conference} on {Software} {Engineering}: {Software} {Engineering} in {Society} ({ICSE}-{SEIS})},
	pages        = {31--40},
	doi          = {10.1109/ICSE-SEIS.2019.00012},
}

@inproceedings{zhang_are_2018,
	title        = {{Are code examples on an online {Q}\&{A} forum reliable? a study of {API} misuse on stack overflow}},
	shorttitle   = {Are code examples on an online {Q}\&{A} forum reliable?},
	author       = {Zhang, Tianyi and Upadhyaya, Ganesha and Reinhardt, Anastasia and Rajan, Hridesh and Kim, Miryung},
	year         = 2018,
	month        = may,
	booktitle    = {Proceedings of the 40th {International} {Conference} on {Software} {Engineering}},
	publisher    = {Association for Computing Machinery},
	address      = {New York, NY, USA},
	series       = {{ICSE} '18},
	pages        = {886--896},
	doi          = {10.1145/3180155.3180260},
	isbn         = {978-1-4503-5638-1},
	url          = {https://doi.org/10.1145/3180155.3180260},
	urldate      = {2022-12-01},
}

@inproceedings{meng_secure_2018,
	title        = {{Secure coding practices in {Java}: challenges and vulnerabilities}},
	shorttitle   = {Secure coding practices in {Java}},
	author       = {Meng, Na and Nagy, Stefan and Yao, Danfeng (Daphne) and Zhuang, Wenjie and Argoty, Gustavo Arango},
	year         = 2018,
	month        = may,
	booktitle    = {Proceedings of the 40th {International} {Conference} on {Software} {Engineering}},
	publisher    = {ACM},
	address      = {Gothenburg Sweden},
	pages        = {372--383},
	doi          = {10.1145/3180155.3180201},
	isbn         = {978-1-4503-5638-1},
	url          = {https://dl.acm.org/doi/10.1145/3180155.3180201},
	urldate      = {2023-04-18},
	language     = {en},
}

@inproceedings{imtiaz_challenges_2019,
	title        = {{Challenges with {Responding} to {Static} {Analysis} {Tool} {Alerts}}},
	author       = {Imtiaz, Nasif and Rahman, Akond and Farhana, Effat and Williams, Laurie},
	year         = 2019,
	month        = may,
	booktitle    = {2019 {IEEE}/{ACM} 16th {International} {Conference} on {Mining} {Software} {Repositories} ({MSR})},
	publisher    = {IEEE},
	address      = {Montreal, QC, Canada},
	pages        = {245--249},
	doi          = {10.1109/MSR.2019.00049},
	isbn         = {978-1-72813-412-3},
	url          = {https://ieeexplore.ieee.org/document/8816730/},
	urldate      = {2023-04-18}
}

@inproceedings{rahman_birds_2019,
	title        = {{A bird's eye view of knowledge needs related to penetration testing}},
	author       = {Rahman, Akond and Williams, Laurie},
	year         = 2019,
	month        = apr,
	booktitle    = {Proceedings of the 6th {Annual} {Symposium} on {Hot} {Topics} in the {Science} of {Security}},
	publisher    = {Association for Computing Machinery},
	address      = {New York, NY, USA},
	series       = {{HotSoS} '19},
	pages        = {1--2},
	doi          = {10.1145/3314058.3317294},
	isbn         = {978-1-4503-7147-6},
	url          = {https://doi.org/10.1145/3314058.3317294},
	urldate      = {2023-01-14},
}

@inproceedings{xu_how_2017,
	title        = {{How {Do} {System} {Administrators} {Resolve} {Access}-{Denied} {Issues} in the {Real} {World}?}},
	author       = {Xu, Tianyin and Naing, Han Min and Lu, Le and Zhou, Yuanyuan},
	year         = 2017,
	month        = may,
	booktitle    = {Proceedings of the 2017 {CHI} {Conference} on {Human} {Factors} in {Computing} {Systems}},
	publisher    = {ACM},
	address      = {Denver Colorado USA},
	pages        = {348--361},
	doi          = {10.1145/3025453.3025999},
	isbn         = {978-1-4503-4655-9},
	url          = {https://dl.acm.org/doi/10.1145/3025453.3025999},
	urldate      = {2023-01-13},
	language     = {en}
}

@inproceedings{abubakar_shard_2021,
	title        = {{{SHARD}: {Fine}-{Grained} {Kernel} {Specialization} with {Context}-{Aware} {Hardening}}},
	shorttitle   = {{SHARD}},
	author       = {Abubakar, Muhammad and Ahmad, Adil and Fonseca, Pedro and Xu, Dongyan},
	year         = 2021,
	booktitle    = {30th {USENIX} {Security} {Symposium}, {USENIX} {Security} 2021, {August} 11-13, 2021},
	publisher    = {USENIX Association},
	pages        = {2435--2452},
	url          = {https://www.usenix.org/conference/usenixsecurity21/presentation/abubakar},
	urldate      = {2022-12-23},
	editor       = {Bailey, Michael and Greenstadt, Rachel},
}

@inproceedings{burihabwa_sgx-fs_2018,
	title        = {{{SGX}-{FS}: {Hardening} a {File} {System} in {User}-{Space} with {Intel} {SGX}}},
	shorttitle   = {{SGX}-{FS}},
	author       = {Burihabwa, Dorian and Felber, Pascal and Mercier, Hugues and Schiavoni, Valerio},
	year         = 2018,
	booktitle    = {2018 {IEEE} {International} {Conference} on {Cloud} {Computing} {Technology} and {Science}, {CloudCom} 2018, {Nicosia}, {Cyprus}, {December} 10-13, 2018},
	publisher    = {IEEE Computer Society},
	pages        = {67--72},
	doi          = {10.1109/CloudCom2018.2018.00027},
}

@inproceedings{sun_security_2018,
	title        = {{Security {Namespace}: {Making} {Linux} {Security} {Frameworks} {Available} to {Containers}}},
	shorttitle   = {Security {Namespace}},
	author       = {Sun, Yuqiong and Safford, David and Zohar, Mimi and Pendarakis, Dimitrios and Gu, Zhongshu and Jaeger, Trent},
	year         = 2018,
	pages        = {1423--1439},
	isbn         = {978-1-939133-04-5},
	url          = {https://www.usenix.org/conference/usenixsecurity18/presentation/sun},
	urldate      = {2022-12-01},
	language     = {en},
}

@inproceedings{wang_running_2019,
	title        = {{Running {Language} {Interpreters} {Inside} {SGX}: {A} {Lightweight}, {Legacy}-{Compatible} {Script} {Code} {Hardening} {Approach}}},
	shorttitle   = {Running {Language} {Interpreters} {Inside} {SGX}},
	author       = {Wang, Huibo and Bauman, Erick and Karande, Vishal and Lin, Zhiqiang and Cheng, Yueqiang and Zhang, Yinqian},
	year         = 2019,
	booktitle    = {Proceedings of the 2019 {ACM} {Asia} {Conference} on {Computer} and {Communications} {Security}, {AsiaCCS} 2019, {Auckland}, {New} {Zealand}, {July} 09-12, 2019},
	publisher    = {ACM},
	pages        = {114--121},
	doi          = {10.1145/3321705.3329848},
	editor       = {Galbraith, Steven D. and Russello, Giovanni and Susilo, Willy and Gollmann, Dieter and Kirda, Engin and Liang, Zhenkai},
}

@inproceedings{wright_linux_2002,
	title        = {{Linux {Security} {Modules}: {General} {Security} {Support} for the {Linux} {Kernel}}},
	shorttitle   = {Linux {Security} {Modules}},
	author       = {Wright, Chris and Cowan, Crispin and Smalley, Stephen and Morris, James and Kroah-Hartman, Greg},
	year         = 2002,
	url          = {https://www.usenix.org/conference/11th-usenix-security-symposium/linux-security-modules-general-security-support-linux},
	urldate      = {2022-12-01},
	language     = {en},
}

@inproceedings{lin_grebe_2022,
	title        = {{{GREBE}: {Unveiling} {Exploitation} {Potential} for {Linux} {Kernel} {Bugs}}},
	shorttitle   = {{GREBE}},
	author       = {Lin, Zhenpeng and Chen, Yueqi and Wu, Yuhang and Mu, Dongliang and Yu, Chensheng and Xing, Xinyu and Li, Kang},
	year         = 2022,
	month        = may,
	booktitle    = {2022 {IEEE} {Symposium} on {Security} and {Privacy} ({SP})},
	pages        = {2078--2095},
	doi          = {10.1109/SP46214.2022.9833683},
	note         = {ISSN: 2375-1207},
}

@inproceedings{tian_lbm_2019,
	title        = {{{LBM}: {A} {Security} {Framework} for {Peripherals} within the {Linux} {Kernel}}},
	shorttitle   = {{LBM}},
	author       = {Tian, Dave Jing and Hernandez, Grant and Choi, Joseph I. and Frost, Vanessa and Johnson, Peter C. and Butler, Kevin R. B.},
	year         = 2019,
	month        = may,
	booktitle    = {2019 {IEEE} {Symposium} on {Security} and {Privacy} ({SP})},
	pages        = {967--984},
	doi          = {10.1109/SP.2019.00041},
	note         = {ISSN: 2375-1207},
}

@inproceedings{tahaei_understanding_2020,
	title        = {{Understanding {Privacy}-{Related} {Questions} on {Stack} {Overflow}}},
	author       = {Tahaei, Mohammad and Vaniea, Kami and Saphra, Naomi},
	year         = 2020,
	month        = apr,
	booktitle    = {Proceedings of the 2020 {CHI} {Conference} on {Human} {Factors} in {Computing} {Systems}},
	publisher    = {ACM},
	address      = {Honolulu HI USA},
	pages        = {1--14},
	doi          = {10.1145/3313831.3376768},
	isbn         = {978-1-4503-6708-0},
	url          = {https://dl.acm.org/doi/10.1145/3313831.3376768},
	urldate      = {2022-12-16},
	language     = {en},
}

@inproceedings{ozga_chors_2022,
	title        = {{{CHORS}: hardening high-assurance security systems with trusted computing}},
	shorttitle   = {{CHORS}},
	author       = {Ozga, Wojciech and Faqeh, Rasha and Quoc, Do Le and Gregor, Franz and Dragone, Silvio and Fetzer, Christof},
	year         = 2022,
	month        = may,
	booktitle    = {Proceedings of the 37th {ACM}/{SIGAPP} {Symposium} on {Applied} {Computing}},
	publisher    = {Association for Computing Machinery},
	address      = {New York, NY, USA},
	series       = {{SAC} '22},
	pages        = {1626--1635},
	doi          = {10.1145/3477314.3506961},
	isbn         = {978-1-4503-8713-2},
	url          = {https://doi.org/10.1145/3477314.3506961},
	urldate      = {2022-12-02},
}

@article{sasidharan_case_2022,
	title        = {{A {Case} {Study} to {Implement} {Windows} {System} {Hardening} using {CIS} {Controls}}},
	author       = {Sasidharan, Rajeshkumar},
	year         = 2022,
	month        = jul,
	journal      = {International Journal of Computer Trends and Technology},
	volume       = 70,
	pages        = {1--7},
	doi          = {10.14445/22312803/IJCTT-V70I7P101},
}

@inproceedings{h_system_2024,
	title        = {{System {Hardening} using {CIS} {Benchmarks}}},
	author       = {H, Ambika P. and Sujatha, G.},
	year         = 2024,
	month        = may,
	booktitle    = {2024 {International} {Conference} on {Advances} in {Computing}, {Communication} and {Applied} {Informatics} ({ACCAI})},
	pages        = {1--6},
	doi          = {10.1109/ACCAI61061.2024.10602274},
	url          = {https://ieeexplore.ieee.org/abstract/document/10602274?casa\%5Ftoken=qKeh-klKnm8AAAAA:65EV9Nx-wGF60mqmGRgegYLAVrFJiUUH4hgtyUmKQ2U3ttHoopsUnN16nX\%5FFJpjhCnj2R1Orkc0},
	urldate      = {2024-09-03},
}

\appendix

\section{\break Stack Exchange Posts}

\begin{table}[htp]
    \centering
    \caption{Number of posts extracted by Stack Exchange platform.}
    \label{tab:qeuscnt}
    \footnotesize
    \renewcommand{\arraystretch}{1.00}
    \setlength{\tabcolsep}{0.75\tabcolsep}
    \setlength{\defaultaddspace}{0.5\defaultaddspace} %
    \rowcolors{2}{white}{gray!10}
    \begin{tabular}{lr}
        \toprule
                            \textbf{Platform} &  \textbf{Posts} \\
        \midrule
                security.stackexchange.com & \postssecurity\\
                stackoverflow.com & \postsstackoverflow\\
                serverfault.com & \postsserverfault\\
                unix.stackexchange.com & \postsunix\\
                askubuntu.com & \postsaskubuntu\\
                sitecore.stackexchange.com & \postssitecore\\
                superuser.com & \postssuperuser\\
                apple.stackexchange.com & \postsapple\\
                raspberrypi.stackexchange.com & \postsraspberrypi\\
                crypto.stackexchange.com & \postscrypto\\
                networkengineering.stackexchange.com & \postsnetworkengineering\\
                dba.stackexchange.com & \postsdba\\
                devops.stackexchange.com & \postsdevops\\
                drupal.stackexchange.com & \postsdrupal\\
                joomla.stackexchange.com & \postsjoomla\\
                wordpress.stackexchange.com & \postswordpress\\
                monero.stackexchange.com & \postsmonero\\
                tor.stackexchange.com & \poststor\\
                ethereum.stackexchange.com & \postsethereum\\
                webmasters.stackexchange.com & \postswebmasters\\
                softwarerecs.stackexchange.com & \postssoftwarerecs\\
                codegolf.stackexchange.com & \postscodegolf\\
                \midrule
                $\sum$ without Duplicates & \var{posts.started}\\
                Duplicates & \var{questions.count.duplicates}\\
                \midrule
                \textbf{Total} & \var{posts.started_with_duplicates}\\
        \bottomrule
    \end{tabular}
\end{table}

\begin{table*}[ht]
        \scriptsize
        \caption{Stack Exchange posts mentioned in this paper.}
        \label{tab:stackexchange}
        \renewcommand{\arraystretch}{1.33}
        \setlength{\tabcolsep}{0.6\tabcolsep}
        \setlength{\defaultaddspace}{0.1\defaultaddspace} %
        \rowcolors{2}{white}{gray!10}
        \begin{tabularx}{\linewidth}{lXl}
            \toprule
            \textbf{ID} & \textbf{Title} & \textbf{URL} \\
            \midrule
            
            83606 & A recommended Tool/Software for hardening Windows Defender & \url{https://softwarerecs.stackexchange.com/questions/83606} \\ 
            113063 & How to disable services on DietPi (debian linux) and harden the security? & \url{https://raspberrypi.stackexchange.com/questions/113063} \\ 
            182508 & Modify Win OS banner to avoid OS detection & \url{https://security.stackexchange.com/questions/182508} \\ 
            191469 & Securing a Laptop from a Foreign Intelligence Agency & \url{https://security.stackexchange.com/questions/191469} \\ 
            195359 & How to restrict plugins/tools like Netcraft and Builtwith to detect server side technologies? & \url{https://security.stackexchange.com/questions/195359} \\ 
            195845 & Where to find the security hardened docker images & \url{https://security.stackexchange.com/questions/195845} \\ 
            199246 & WIN 10 hardening: Importance of “Allow system to be shut down without having to log on” policy & \url{https://security.stackexchange.com/questions/199246} \\ 
            204026 & CIS hardening of alpine based docker container & \url{https://security.stackexchange.com/questions/204026} \\ 
            210589 & Why is autofs insecure? & \url{https://security.stackexchange.com/questions/210589} \\ 
            215398 & How should you configure PowerShell logs permissions? & \url{https://security.stackexchange.com/questions/215398} \\ 
            219981 & Tightly locking down a FreeBSD jail & \url{https://security.stackexchange.com/questions/219981} \\ 
            222616 & Restrict privileged users from accessing certain directories on Linux servers with Grsecurity? & \url{https://security.stackexchange.com/questions/222616} \\ 
            231046 & My WP site just got hacked for the third time even after following WP hardening guidelines & \url{https://security.stackexchange.com/questions/231046} \\ 
            406843 & Extreme hardening of my Macbook pro from physical hacking & \url{https://apple.stackexchange.com/questions/406843} \\ 
            487876 & Is it useful to set the policies to DROP for all tables in Iptables? & \url{https://unix.stackexchange.com/questions/487876} \\ 
            645077 & hidepid=2 stopped working after an update. Kernel don’t suppport “per-mount point”? & \url{https://unix.stackexchange.com/questions/645077} \\ 
            899084 & Solaris 11 Auditing, audit\_control file cannot be found & \url{https://serverfault.com/questions/899084} \\ 
            941192 & Windows 10: Kerberos settings not found & \url{https://serverfault.com/questions/941192} \\ 
            955208 & Attempting to delete cron directory in docker gives “Invalid argument” & \url{https://serverfault.com/questions/955208} \\ 
            969212 & Apache HTTPD and Tomcat - how to harden and/or disable Apache when only using Tomcat? & \url{https://serverfault.com/questions/969212} \\ 
            1018828 & Editing Authselect files & \url{https://serverfault.com/questions/1018828} \\ 
            1027188 & iptables policy \& saving in RHEL & \url{https://serverfault.com/questions/1027188} \\ 
            1090794 & Network Security: Hardening IPv6 on Ubuntu Server? & \url{https://serverfault.com/questions/1090794} \\ 
            1361197 & Accounts Expired after CIS Hardening on Ubuntu 20.04 - Workstation Level 1 & \url{https://askubuntu.com/questions/1361197} \\ 
            1460813 & gnome-terminal/shell hardening security & \url{https://askubuntu.com/questions/1460813} \\ 
            1720993 & Network share access denied after STIG/CIS hardening in windows & \url{https://superuser.com/questions/1720993} \\ 
            54459236 & Wordpress Default Directory Change & \url{https://stackoverflow.com/questions/54459236} \\ 
            56143561 & How to Harden Apache against security vulnerabilities & \url{https://stackoverflow.com/questions/56143561} \\
            57731428 & How do I prevent root access to my docker container & \url{https://stackoverflow.com/questions/57731428} \\ 
            58132270 & Will Memory Tagging Extension be implemented in x86? & \url{https://stackoverflow.com/questions/58132270} \\ 
            68112625 & CIS hardening script for windows 2016 server in GCP & \url{https://stackoverflow.com/questions/68112625} \\ 
            69471721 & How do you use the win security policy module for something in the local policies section using Ansible? & \url{https://stackoverflow.com/questions/69471721} \\ 
            70463445 & Windows 2019 CIS benchmark audit tool/script & \url{https://stackoverflow.com/questions/70463445} \\ 
            73808977 & Daemon Umask in RHEL 8 & \url{https://stackoverflow.com/questions/73808977} \\ 
            75371459 & Oracle Linux 8 hardening with CIS security policy & \url{https://stackoverflow.com/questions/75371459} \\
            \bottomrule
        \end{tabularx}
\end{table*}

\end{document}